\documentclass[12pt]{article}

\usepackage{amsfonts,amsmath,amssymb}
\usepackage{cite}
\usepackage{epsfig}
\usepackage{paralist}
\usepackage{fancyhdr}
\usepackage{tikz}
\usepackage{tkz-euclide}
\usetikzlibrary{decorations.pathmorphing,patterns,calc,snakes,arrows}

\usepackage{graphicx}
\usepackage{xcolor}
\numberwithin{equation}{section}
\usepackage[vcentermath]{youngtab}
\usepackage{etex}
\usepackage{braket}
\usepackage{float}
\usepackage{placeins}

\def\spa#1{\phantom{\fbox{\rule[-#1cm]{0cm}{0cm}}}}

\def\be{\begin{equation}}
\def\ee{\end{equation}}
\def\bea{\begin{eqnarray}}
\def\eea{\end{eqnarray}}

\def\del{\partial}

\renewcommand{\thefootnote}{\fnsymbol{footnote}}

\makeatletter
\g@addto@macro\bfseries{\boldmath}
\makeatother

\def\Tb{{\bar{T}}}

\usepackage{hyperref}
\begin{document}

\hfuzz=100pt
\title{{\Large \bf{Wavefunctions of AdS$_3$ Universes and $T\Tb$-deformed Torus Partition Functions}}}
\date{}
\author{Shinji Hirano$^{a, b}$\footnote{
	e-mail:
	\href{mailto:shinji.hirano@gmail.com}{shinji.hirano@gmail.com}}}
\date{}

\maketitle

\thispagestyle{fancy}
\rhead{YITP-26-103}
\cfoot{}
\renewcommand{\headrulewidth}{0.0pt}

\vspace*{-1cm}
\begin{center}
$^{a}${{\it School of Science, Huzhou Normal University}}
\\ {{\it Huzhou 313000, Zhejiang, China}}
  \spa{0.5} \\
$^b${{\it Center for Gravitational Physics and Quantum Information (CGPQI)}}
\\ {{\it  Yukawa Institute for Theoretical Physics, Kyoto University}}
\\ {{\it Kitashirakawa-Oiwakecho, Sakyo-ku, Kyoto 606-8502, Japan}}
\spa{0.5}  

\end{center}

\begin{abstract}
We study wavefunctions of quantum gravity in asymptotically AdS$_3$ spacetimes and their relation to $T\bar T$-deformed torus partition functions. We show that the deformed partition function is given by an invertible integral transform of bulk wavefunctions, which implements a change of representation from the York time to its conjugate, the spatial volume. The invertibility of the transform allows the bulk wavefunction to be reconstructed from the family of $T\bar T$-deformed partition functions.
A $T\bar T$-deformed partition function at fixed coupling is the amplitude of the bulk state in a finite-width, fixed-volume wavepacket localized at the asymptotic boundary in York time. In the CFT limit, the wavepacket localizes at the boundary, recovering the standard holographic relation. The holographic screen of a given state is the support of the integrand of the transform: at large central charge, it separates from the wavepacket and reproduces finite-cutoff holography for the sign of the coupling with complex energies, and glue-on holography for the opposite sign.
In Euclidean signature, the construction arises naturally from wavefunctions of Rindler AdS$_3$, while after a double Wick rotation it admits a Lorentzian interpretation in terms of closed AdS$_3$ torus universes without asymptotic boundaries. We also discuss de Sitter torus universes and generalizations to other spatial topologies and higher dimensions.

\end{abstract}

\renewcommand{\thefootnote}{\arabic{footnote}}
\setcounter{footnote}{0}

\newpage

\tableofcontents


\section{Introduction}
\label{sec:Introduction}

One of the fundamental questions in holography is how bulk spacetime is reconstructed from boundary data. While the AdS/CFT correspondence \cite{Maldacena:1997re} identifies the boundary partition function with the Wheeler--DeWitt (WdW) wavefunction evaluated at the asymptotic boundary of AdS \cite{Witten:1998qj, Gubser:1998bc}, canonical quantum gravity naturally describes its radial evolution away from the asymptotic boundary through the WdW equation. The relation between bulk WdW wavefunctions and boundary partition functions has also been investigated from the canonical perspective \cite{Freidel:2008sh}. How the bulk wavefunction is encoded in the boundary theory, however, remains much less understood.

An important clue to this question emerged from the discovery of the
$T\bar T$ deformation
\cite{Zamolodchikov:2004ce,Smirnov:2016lqw,Cavaglia:2016oda}. A major
breakthrough came with the proposal that the $T\bar T$ deformation is
holographically dual to a finite radial cutoff in AdS$_3$
\cite{McGough:2016lol}. This proposal was motivated by the close
relation between the $T\bar T$ flow equation and the 
Hamiltonian constraint of three-dimensional gravity. It thereby
established a concrete connection between $T\bar T$ deformations and the
canonical description of bulk radial evolution. This connection was
subsequently developed further in~\cite{Kraus:2018xrn, Caputa:2020lpa},
reinforcing the view that the $T\bar T$ deformation provides a boundary
description of bulk evolution. 
Building on these developments, Witten reformulated the
$T\bar T$/WdW correspondence within the framework of canonical
quantization \cite{Witten:2022xxp}. More generally, Wall and
collaborators advocated a canonical formulation of holography, motivated
in part by the relation to $T^2$ deformations, in which the WdW
wavefunction plays the central role and holographic data may be
associated with arbitrary Cauchy slices
\cite{Araujo-Regado:2022gvw}. Taken together, these developments suggest a close connection between
$T\bar T$-deformed partition functions and the WdW wavefunction, while
the precise holographic realization of this correspondence has remained
unclear.

In this paper, we show that the relation between the $T\bar T$-deformed
partition function and the WdW wavefunction is more subtle than a direct
identification on a sharply defined radial slice. Instead, we show in
the case of the torus that the $T\bar T$-deformed partition function is
related to the WdW wavefunction by an explicit invertible integral
transform, which implements a change of representation from the York time to its conjugate, the spatial volume. In the CFT limit, the kernel
localizes at the asymptotic boundary, recovering the standard AdS/CFT
relation. At finite deformation, a $T\bar T$-deformed torus partition function at fixed
coupling is the bulk wavefunction at fixed volume, whose radial position is therefore not sharp: it is naturally interpreted as a bulk wavepacket of finite radial extent.\footnote{For an earlier study of a closely related idea, see~\cite{Coleman:2020jte}. We comment further on its relation to the present work in footnote~\ref{Coleman_Shyam}.}

The kernel transform has several important consequences. Its invertibility implies that the bulk WdW wavefunction can be reconstructed from the family of $T\bar T$-deformed partition functions, thereby establishing the precise holographic relation between the boundary partition functions and the bulk quantum state. It applies to both signs of the coupling: for the sign with Hagedorn growth, it is the projection of the Lorentzian Schr\"odinger wavefunction onto volume eigenstates, while for the sign with complex energies it runs through the classically forbidden window of York time, and the positive-frequency condition of the reduced phase space fixes the nonperturbative ambiguity left by the CFT limit.

The transform also refines finite-cutoff holography. At large central charge, the semiclassical part of the bulk state localizes on a finite cutoff surface whose radius and quasi-local energy reproduce the relation of \cite{McGough:2016lol,Kraus:2018xrn} for the sign with complex energies. Finite-cutoff holography is thus derived at leading order, with a cutoff surface of width of order $c^{-1/2}$. For the opposite sign, the semiclassical screen lies instead on a glue-on surface beyond the asymptotic boundary~\cite{Apolo:2023vnm}, and both cases are captured by the single relation $r_c^2=-4G\ell/(\pi\mu)$. The screen interpretation thus differs sharply between the two signs: it is the bulk state, rather than the kernel, that determines on which side of the boundary the screen lies. The normalizable part of the bulk state defines no screen: its perturbative corrections arise from complex geometries in the classically forbidden window, and, for the sign with complex energies, its nonperturbative part from slices behind the horizon, a bulk counterpart of the complex spectrum. After a double Wick rotation, the construction admits a Lorentzian interpretation in terms of York-time evolution, leading naturally to a holographic description of closed AdS$_3$ torus universes without asymptotic boundaries.

Existing approaches, such as HKLL \cite{Hamilton:2006az} and entanglement wedge reconstruction \cite{Czech:2012bh, Headrick:2014cta, Wall:2012uf, Jafferis:2015del, Dong:2016eik, Cotler:2017erl}, focus on reconstructing local bulk operators or operator algebras from boundary observables. By contrast, we reconstruct the bulk WdW wavefunction, providing a holographic description of the quantum state itself. These complementary perspectives address different aspects of bulk reconstruction and together contribute to a more complete understanding of holography.

The remainder of this paper is organized as follows.
In Section~2, we review the reduced phase-space quantization of AdS$_3$ torus universes.
In Section~3, we derive the integral transform relating the $T\bar T$-deformed torus partition function to the bulk WdW wavefunction and develop its holographic interpretation.
Section~4 presents the inverse transform and reconstructs the bulk wavefunction from the family of $T\bar T$-deformed partition functions.
Section~5 extends the construction to de Sitter torus universes and discusses its implications for de Sitter holography.
Section~6 discusses possible generalizations of the present framework to
more general boundary geometries and higher-dimensional spacetimes.
Finally, Section~7 concludes by summarizing the implications of the
present work for holographic bulk reconstruction and the interpretation
of the holographic screen in Lorentzian AdS and de Sitter spacetimes.

\section{Canonical Quantization of AdS$_3$ Torus Universes}
\label{sec:CQTUAdS}

As discussed in the introduction, the central objects of our study are the wavefunctions of asymptotically AdS$_3$ universes. We focus on the tractable yet nontrivial case of torus universes to develop an explicit correspondence with the $T\bar{T}$-deformed partition function. We begin by reviewing their canonical quantization in the reduced phase-space formulation \cite{Moncrief:1989dx, Hosoya:1989yj, Fujiwara:1989xg, Ezawa:1993ti, Carlip:2004ba, Carlip:1991ij, Carlip:1994ap, Carlip:1992cj}.

To establish our notation, we adopt the ADM decomposition of the spacetime metric,
\begin{align}\label{ADM}
ds^2=-N^2dt^2+g_{ij}\left(dx^i+N^i dt\right)\left(dx^j+N^j dt\right)\,.
\end{align}
We work in the constant-mean-curvature (CMC), or York, gauge \cite{York1973}, in which the trace of the extrinsic curvature is constant on each Cauchy slice,
\begin{align}
K=-\tau(t),
\end{align}
where \(\tau(t)\) serves as the York time parametrizing the foliation. In this gauge, the momentum constraint,
\begin{align}\label{momentum_constraints}
\nabla_i\left(\frac{\pi^{ij}}{\sqrt{g}}\right)=0\ ,
\qquad
\pi^{ij}=\sqrt{g}\left(K^{ij}-g^{ij}K\right)\ ,
\end{align}
is equivalent to 
\begin{align}\label{MC_York}
0=\nabla_i\left(K^{ij}-\frac{1}{d-1}g^{ij}K\right)
\equiv
\nabla_i\Sigma^{ij}\ ,
\end{align}
where \(\nabla_i\) is the covariant derivative associated with the spatial metric \(g_{ij}\). Since \(K\) is constant on each Cauchy slice, its gradient vanishes and therefore drops out of the momentum constraint. Consequently, \(\Sigma^{ij}\) is both traceless and transverse. 

The key idea of the reduced phase-space formulation is to decompose the spatial metric as
\begin{align}
g_{ij}=e^{2\phi}\bar{g}_{ij},
\end{align}
where \(\phi\) is the conformal factor, while fixing the topology of the spatial manifold. One then introduces the conformally rescaled traceless tensor
\begin{align}
\bar{\Sigma}^{ij}=e^{(d+1)\phi}\Sigma^{ij},
\end{align}
whose covariant components satisfy
\begin{align}
\bar{\Sigma}_{ij}
=\bar{g}_{ik}\bar{g}_{jl}\bar{\Sigma}^{kl}
=e^{(d-3)\phi}\Sigma_{ij}.
\end{align}
The momentum constraint \eqref{MC_York} then implies that \(\bar{\Sigma}^{ij}\) is transverse with respect to the freely chosen reference metric \(\bar g_{ij}\),
\begin{align}\label{quadratic_diff}
\bar{\nabla}_i\bar{\Sigma}^{ij}=0,
\end{align}
where \(\bar{\nabla}_i\) denotes the covariant derivative associated with \(\bar g_{ij}\). In particular, in \(d=3\) spacetimes, \(\bar{\Sigma}^{ij}\) is naturally identified with a holomorphic quadratic differential on the spatial Riemann surface.

The Hamiltonian constraint determines the conformal factor $\phi$
through the Lichnerowicz equation,
\begin{equation}
\hspace{-.3cm}
\begin{aligned}\label{L_eqn}
0&={\cal H}
=K_{ij}K^{ij}-K^2-R+2\Lambda\\
&=e^{2(1-d)\phi}\bar{\Sigma}_{ij}\bar{\Sigma}^{ij}
-\frac{d-2}{d-1}\tau^2
-e^{-2\phi}\!\left(
\bar{R}
-2(d-2)\bar{\nabla}^2\phi
-(d-2)(d-3)\bar{\nabla}^i\phi\bar{\nabla}_i\phi
\right)
+2\Lambda .
\end{aligned}
\end{equation}
Under suitable conditions, this equation admits a unique positive solution
for the conformal factor. The momentum constraint is solved by
constructing transverse-traceless (TT) tensors $\bar{\Sigma}^{ij}$ on a chosen reference
geometry, while the Hamiltonian constraint fixes the remaining conformal
degree of freedom. Consequently, all constraints are solved, and the
reduced phase space is parametrized by the conformal geometry of the
spatial slice together with its conjugate TT momentum
\cite{Fischer:1996qg}. In dimensions $d>3$, these variables describe the
local propagating graviton degrees of freedom. In $d=3$, where there are
no local graviton modes, the reduced phase space reduces to the
cotangent bundle of the Teichm\"uller space of the spatial manifold
\cite{Moncrief:1989dx}.

To study the relation between bulk reconstruction and the $T\bar{T}$ deformation, 
we now turn to the simplest example: torus universes in asymptotically (A)dS$_3$ spacetime. 
For the flat torus, $\bar{R}=0$ and $\bar{\Sigma}^2\equiv \bar{\Sigma}^{ij}\bar{\Sigma}_{ij}$ is constant. 
The unique solution for the conformal factor is spatially constant, and the Lichnerowicz equation is solved by
\begin{align}\label{H_solved}
e^{2\phi}=\sqrt{2\bar{\Sigma}^2}(\tau^2-4\Lambda)^{-1/2}\ .
\end{align}
With both the Hamiltonian and momentum constraints solved, the Einstein--Hilbert action reduces to
\begin{align}
S_{EH}=\int dt\int_{{\cal M}_2} d^2x\,\pi^{ij}\dot{g}_{ij}\ ,
\end{align}
where $\pi^{ij}$ is the canonical momentum defined in \eqref{momentum_constraints}. This can be rewritten as
\begin{equation}
\begin{aligned}\label{EH_ADM}
S_{EH}
&=\int dt\int_{{\cal M}_2}d^2x
\left(
e^{2\phi}\sqrt{g}\,\Sigma^{ij}\dot{\bar{g}}_{ij}
+\tau\,\partial_t\sqrt{g}
\right)\\
&=\int d\tau
\left(
p^a\partial_{\tau}m_a-V
\right),
\end{aligned}
\end{equation}
where the momentum $p^a$ conjugate to the moduli $m_a$ and the spatial volume $V$ are defined by
\begin{align}\label{pa_V}
p^a
&\equiv
\int_{{\cal M}_2}d^2x\,
\sqrt{\bar g}\,
\bar{\Sigma}^{ij}
\frac{\partial\bar g_{ij}}{\partial m_a}\ ,\qquad\qquad
V\equiv
\int_{{\cal M}_2}d^2x\,\sqrt{g}\ .
\end{align}
Here $m_a$ denote the moduli parametrizing the conformal geometry of the Cauchy slice. The final expression in \eqref{EH_ADM} defines the reduced Hamiltonian system, in which $(m_a,p^a)$ form canonical conjugate pairs, the Hamiltonian is
\begin{align}
H_{\rm red}=V=e^{2\phi}\,\bar V(m_a),
\end{align}
and York time $\tau$ serves as the evolution parameter.

For the torus, 
\begin{align}
d\bar{s}^2
=\bar{g}_{ij}dx^idx^j
=\frac{1}{m_2}|dx+mdy|^2\ ,
\qquad
0\le x,y\le1,
\end{align}
the reference volume is $\bar V=1$. Hence,
\begin{align}
V=e^{2\phi}
=\sqrt{2\bar{\Sigma}^2}\,
(\tau^2-4\Lambda)^{-1/2}.
\end{align}
Introducing the complex coordinate
\begin{align}
z=x+my,
\end{align}
one finds
\begin{align}
p^1
&=2i\left(\bar{\Sigma}_{zz}-\bar{\Sigma}_{\bar z\bar z}\right)\ ,\qquad
p^2
=-2\left(\bar{\Sigma}_{zz}+\bar{\Sigma}_{\bar z\bar z}\right)\ ,
\end{align}
which implies
\begin{align}
(p^1)^2+(p^2)^2
=
16\,\bar{\Sigma}_{zz}\bar{\Sigma}_{\bar z\bar z}
=
\frac{2\bar{\Sigma}^2}{m_2^2}\ .
\end{align}
Therefore, the Hamiltonian of the reduced system is
\begin{align}
H_{\rm red}=V=\frac{\sqrt{m_2^2\,p\bar p}}{\sqrt{\tau^2-4\Lambda}}\ ,
\qquad\mbox{where}\qquad
p=p^1+ip^2.
\end{align}
Upon quantization, this becomes, up to operator-ordering ambiguities, the Schr\"odinger equation \cite{Carlip:1994ap}
\begin{align}\label{Schrodinger}
i\hbar\frac{\partial}{\partial\tau}\Psi(\tau,m_a)
=
\frac{1}{\sqrt{\tau^2-4\Lambda}}
\sqrt{m_2^2\,p\bar p}\,
\Psi(\tau,m_a)\ ,
\end{align}
where the canonical variables satisfy the commutation relations
\begin{align}
[m_a,p^b]=i\hbar\,\delta_a^{\,b}\ .
\end{align}
It is instructive to compare the Schr\"odinger equation \eqref{Schrodinger}
with the Wheeler--DeWitt (WdW) equation \cite{Carlip:1991ij}. The Hamiltonian constraint \eqref{L_eqn} can be written as\footnote{\label{WdWEq1}Equation \eqref{H_solved} holds only after solving the Hamiltonian constraint classically, whereas here the constraint itself is being promoted to an operator equation. The rewriting therefore relies solely on the definitions in \eqref{pa_V}.}
\begin{align}\label{Comparison_WdW_Sch}
0={\cal H}
=e^{-4\phi}\bar{\Sigma}^2-{1\over 2}\tau^2+2\Lambda=\frac{1}{2}V^{-2}
\left[
m_2^2p\bar p-(\tau^2-4\Lambda)V^2
\right]\ .
\end{align}
The second term in the first line of \eqref{EH_ADM} shows that $(V,\tau)$
form a canonical conjugate pair, satisfying
\begin{align}
[V,\tau]=i\hbar\ ,
\end{align}
so that $V$ generates York-time translations, as in \eqref{Schrodinger}.
Upon quantization, the Hamiltonian constraint therefore becomes, up to
operator-ordering ambiguities,
\begin{align}\label{WdW}
\left[
m_2^2p\bar p+\hbar^2(\tau^2-4\Lambda)
\frac{\partial^2}{\partial\tau^2}
\right]
\Psi(\tau,m_a)
=0\ .
\end{align}
With an appropriate choice of operator ordering, the WdW
equation is simply the square of the Schr\"odinger equation \eqref{Schrodinger}.

In this paper, we adopt the following operator ordering: First, the relation between the Schr\"odinger and WdW equations uniquely fixes the ordering of $(\tau,V)$ in the Hamiltonian constraint according to
\begin{align}
(\tau^2-4\Lambda)V^2
\;\longrightarrow\;
\left(\sqrt{\tau^2-4\Lambda}\,V\right)^2\ .
\end{align}
The ordering of $(m_a,p^a)$ is chosen as
\begin{align}
m_2^2(p^2)^2
\;\longrightarrow\;
m_2D_y^2m_2^{-1}\ ,
\qquad
D_y\equiv\frac{1}{2}\left(m_2p^2+p^2m_2\right)\ ,
\end{align}
for reasons that will become apparent later. With this choice, the kinetic operator is naturally identified with the shifted Maass Laplacian
\begin{align}
\hspace{-.0cm}
m_2^2p\bar p
\;\longrightarrow\; m_2^2(p^1)^2+m_2D_y^2m_2^{-1}=-\hbar^2\left(\Delta_{\rm Maass}+\frac14\right) \equiv -\hbar^2\left(m_2^2(\partial_1^2+\partial_2^2)+\frac14\right)\ .
\end{align}
The appearance of the universal shift by $1/4$ will play an important role in the subsequent analysis. 
With the above operator ordering, the WdW equation takes the remarkably simple form
\begin{align}\label{L_WdW}
\left(\sqrt{\tau^2-4\Lambda}\,{\partial\over\partial\tau}\right)^2\Psi(\tau, m_a)=\left(\Delta_{\rm Maass}+\frac14\right) \Psi(\tau, m_a)\ .
\end{align}
In the next section, we establish the relation between this equation and the $T\bar{T}$ flow equation for the torus partition function through a nontrivial integral transform.

Before concluding this review, we briefly discuss the Euclidean continuation of the canonical formalism in the radial ADM foliation, which will also be needed in our later analysis of Euclidean AdS$_3$ with a torus boundary. This continuation should not be confused with the double Wick rotation introduced later: here the analytic continuation is implemented by rotating the lapse function, $N\to iN$, while keeping the radial foliation fixed, whereas the latter acts on the spacetime coordinates.

Under the continuation $N\to iN$, the extrinsic curvature transforms as
\begin{align}
K_{ij}\to -iK_{ij},
\end{align}
so that the Hamiltonian constraint becomes
\begin{align}
0={\cal H}_E
=
-K_{ij}K^{ij}
+K^2
-R
+2\Lambda\ ,
\end{align}
where the Lorentzian combination $K_{ij}K^{ij}-K^2$ is replaced by its negative.
The discussion above carries over straightforwardly in the Euclidean setting. Denoting the (real) Euclidean York time by $\tau_E$, the solution to the Lichnerowicz equation in the CMC (York) gauge becomes
\begin{align}\label{H_solved_E}
e^{2\phi}
=
\sqrt{2\bar{\Sigma}_E^2}
(\tau_E^2+4\Lambda)^{-1/2}\ ,
\end{align}
where $\bar\Sigma_E^2\ge0$ is built from the Euclidean extrinsic curvature; in terms of the Lorentzian variables, $\bar\Sigma^2=-\bar\Sigma_E^2$.
Proceeding as in the Lorentzian case, one arrives at the Euclidean WdW equation
\begin{align}\label{E_WdW}
\left(
\sqrt{\tau_E^2+4\Lambda}\,
\frac{\partial}{\partial\tau_E}
\right)^2
\Psi(\tau_E,m_a)
=
\left(
\Delta_{\rm Maass}
+\frac14
\right)
\Psi(\tau_E,m_a)\ .
\end{align}
Although \eqref{E_WdW} takes the same form as \eqref{L_WdW}, the continuation $N\to iN$ also rotates the momenta, $p^a\to-ip^a_E$, so that $p\bar p=(p^1)^2+(p^2)^2$ becomes negative for a real Euclidean geometry, keeping $V$ real as in \eqref{H_solved_E}. With $(\Delta_{\rm Maass}+\frac14)\Psi=-r^2\Psi$, real Euclidean geometries thus correspond to imaginary $r$, outside the square-integrable spectrum, whereas square-integrable modes with real $r$ describe Lorentzian geometries and, in Euclidean York time, complex geometries with imaginary volume. This distinction plays an important role in Section~\ref{sec:Kernel}.

\section{From AdS$_3$ Wavefunctions to $T\bar{T}$-Deformed Torus Partition Functions}
\label{sec:MapWFUandTTbar}

As discussed in the Introduction, the relation between the $T\bar{T}$-deformed torus partition function and the bulk WdW wavefunction is more subtle than a direct identification. In this section, we show that the partition function is related to the bulk wavefunction by an explicit invertible integral transform. The kernel encodes the relation between the $T\bar{T}$ deformation parameter and the bulk radial (or York-time) coordinate, thereby providing a precise holographic dictionary between the boundary theory and the bulk wavefunction.\footnote{The $T\bar{T}$-deformed torus partition function has also been discussed in relation to the WdW wavefunction in the dS$_3$ context in \cite{Godet:2024ich}. The relation derived here is conceptually different: we obtain an explicit invertible integral transform relating the bulk wavefunction to the family of $T\bar{T}$-deformed torus partition functions.}

As discussed in \cite{Datta:2018thy, Aharony:2018bad, Gu:2025tpy, Gu:2026rck}, the $T\bar{T}$-deformed torus partition function satisfies the flow equation
\begin{align}\label{TTbar_flow_mu}
\partial_{\mu}Z_{T\bar{T}}(\mu, m_a)
=
\left(
\frac{1}{4}m_2(\partial_1^2+\partial_2^2)
+\frac{1}{2}\left(\partial_2-\frac{1}{m_2}\right)\mu\partial_{\mu}
\right)
Z_{T\bar{T}}(\mu, m_a)\ ,
\end{align}
where $\mu$ is the dimensionless $T\bar{T}$ coupling, $\mu=2t/(\pi R^2)$, with $t$, of dimensions of length squared, the coupling of the deformation $\partial_t\mathcal{L}=\det T_{\mu\nu}$, and $R$ the radius of the spatial circle on which the energies below are defined \cite{Gu:2025tpy}. The partition function is invariant under modular transformations,
\[
m\rightarrow \frac{am+b}{cm+d}\ ,
\qquad\quad
\mu\rightarrow\frac{\mu}{|cm+d|^2}\ ,
\]
with $m=m_1+im_2$ and $ad-bc=1$ \cite{Aharony:2018bad}.

The two signs of the coupling define distinct theories. For a single state, $Z_{T\bar T}\supset e^{-2\pi m_2E(\mu)+2\pi i m_1P}$, \eqref{TTbar_flow_mu} gives the deformed spectrum
\begin{align}\label{deformed_spectrum}
E(\mu)=\frac{\sqrt{1+2\pi\mu E+\pi^2\mu^2P^2}-1}{\pi\mu}\ ,
\end{align}
where $E$ and $P$ are the undeformed energy and momentum on the spatial circle in units of $1/R$, i.e.\ with $R=1$, so that, e.g., $E=h+\bar h-c/12$ and $P=h-\bar h$. For $\mu>0$ the spectrum stays real and exhibits Hagedorn growth, whereas for $\mu<0$ states above an energy of order $1/|\mu|$ acquire complex energies. The latter sign is associated with a finite radial cutoff in AdS$_3$ \cite{McGough:2016lol,Kraus:2018xrn}, the former with glue-on AdS$_3$ \cite{Apolo:2023vnm}. We call them the complex-energy and Hagedorn branches, and treat both.

To facilitate comparison with the WdW equation, we introduce the modular-invariant deformation parameter and the sign of the coupling,
\[
\lambda\equiv \frac{m_2}{|\mu|}\ ,\qquad\quad \epsilon\equiv{\rm sign}\,\mu\ ,
\]
in terms of which the flow equation becomes
\begin{align}\label{TTbar_flow}
\lambda^2(\partial_{\lambda}^2-4\epsilon\,\partial_{\lambda})
Z_{T\bar{T}}(\lambda,m_a)
=
\Delta_{\rm Maass}
Z_{T\bar{T}}(\lambda,m_a)\ .
\end{align}
This change of variables disentangles the dependence on the modular parameter and the deformation coupling, bringing the flow equation into a form that closely parallels the WdW equation. It is further convenient to redefine
\[
Z_{T\bar{T}}
=
e^{2\epsilon\lambda}\lambda^{1/2}\widetilde{Z}_{T\bar{T}}\ ,
\]
under which the flow equation takes the same form on both branches,\footnote{\label{Coleman_Shyam}
It is worth commenting on the closely related earlier work~\cite{Coleman:2020jte}. As pointed out there, the flow equation for $\lambda^{1/2}\widetilde{Z}_{T\Tb}(\lambda,m_a)$ takes the same form as the Euclidean continuation of the WdW equation~\eqref{Comparison_WdW_Sch} expressed in terms of $V$, provided one adopts a particular operator ordering for the conjugate pair $(\tau,V)$ and identifies $\ell\lambda=V$.
However, this ordering differs from the one obtained in our approach,
which follows directly from the Schr\"odinger equation~\eqref{Schrodinger}
on the (physical) reduced phase space, where $V$ is identified with the
reduced Hamiltonian $H_{\rm red}$. Consequently, even for a wavefunction
at fixed volume, the WdW equation does not coincide with the flow
equation for $\lambda^{1/2}\widetilde{Z}_{T\Tb}(\lambda,m_a)$.
This difference corresponds to a nontrivial Jacobian factor in the Laplace transform introduced in~\cite{Coleman:2020jte}, which appears explicitly in \eqref{Laplace_form} below.}
\begin{align}\label{TTbar_flow_WdW}
\left(
\lambda^2\partial_{\lambda}^2
+\lambda\partial_{\lambda}
-4\lambda^2
\right)
\widetilde{Z}_{T\bar{T}}(\lambda,m_a)
=
\left(
\Delta_{\rm Maass}
+\frac{1}{4}
\right)
\widetilde{Z}_{T\bar{T}}(\lambda,m_a)\ .
\end{align}
This is to be compared with the WdW equation
\begin{align}\label{WdW_T}
\del_T^2
\Psi(T,m_a)
=
\left(
\Delta_{\rm Maass}
+\frac14
\right)
\Psi(T,m_a)\ ,
\end{align}
where the ``time'' $T$ is defined by
\begin{align}\label{define_T}
dT={d\tau\over\sqrt{\tau^2-4\Lambda}}\quad\mbox{for Lorentzian}\ ,\qquad
dT={d\tau_E\over\sqrt{\tau_E^2+4\Lambda}}\quad\mbox{for Euclidean}\ .
\end{align}
The variable \(T\) is a monotonic function of the conventional York time
\(\tau\) (or \(\tau_E\)). Since it plays the role of the evolution parameter
in the WdW equation, we shall refer to it as the {\it WdW time} throughout
this paper.

The common appearance of the shifted Maass Laplacian naturally suggests a spectral decomposition in terms of its eigenfunctions,
\begin{align}
\left(
\Delta_{\rm Maass}
+\frac14
\right)
\phi_r(m_a)
=
-r^2\phi_r(m_a)\ .
\end{align}
The torus partition function itself, however, is not square-integrable on the fundamental domain: it grows exponentially at the cusp because of the vacuum and other light states. We therefore decompose both the $T\bar T$-deformed partition function and the WdW wavefunction into two parts,
\begin{align}\label{two_parts}
F(\xi,m_a)=F_{\rm light}(\xi,m_a)+F_{L^2}(\xi,m_a)\ ,
\end{align}
where $F$ denotes either $\widetilde Z_{T\bar{T}}(\lambda,m_a)$ or $\Psi(T,m_a)$, with $\xi=\lambda$ or $T$, respectively. The light part $F_{\rm light}$ is the modular completion of the contributions of states with $\min(h,\bar h)<(c-1)/24$ and carries the entire growth at the cusp \cite{Benjamin:2021ygh}; it is not a superposition of eigenfunctions of $\Delta_{\rm Maass}$. The remainder $F_{L^2}$ is square-integrable and decomposes into the constant, the Eisenstein series, and the Maass cusp forms,
\begin{align}\label{spectral}
F_{L^2}(\xi,m_a)
=
f_0(\xi)
+\int_0^\infty dr\,f_r(\xi)\,\phi_r^{\rm(E)}(m_a)
+
\sum_n
f_n(\xi)\,
\phi_n^{\rm(cusp)}(m_a)\ ,
\end{align}
where the constant has $r=\pm i/2$ and all other components have real $r$. The same split underlies the third-quantized description of \cite{Hirano:2026vcu}, in which $F_{\rm light}$ is fixed non-normalizable background data and only $F_{L^2}$ spans the one-universe Hilbert space. Here $m_a=(m_1,m_2)$, or equivalently $m=m_1+im_2$, is taken to lie in the fundamental domain
$\mathcal{F}=SL(2,\mathbb{Z})\backslash\mathbb{H}$.

Substituting \eqref{spectral} into Eq.~\eqref{TTbar_flow_WdW}, one finds that each continuous and discrete spectral component (with $r$ replaced by $r_n$ for the latter) satisfies
\begin{align}\label{Bessel}
\left(
\lambda^2\partial_\lambda^2
+\lambda\partial_\lambda
-4\lambda^2
+r^2
\right)f_r(\lambda)=0\ ,
\end{align}
which is the modified Bessel equation. Two conditions select the physical solution. On the bulk side, the reduced phase space selects solutions of the Schr\"odinger equation \eqref{Schrodinger}, which for each spectral component reduces to $i\partial_T\Psi=r\Psi$, rather than of the WdW equation \eqref{WdW_T}. The physical modes are thus of positive frequency,
\begin{align}\label{positive_frequency}
\Psi_r(T)\propto e^{-irT}\ .
\end{align}
On the boundary side, the CFT limit $\lambda\to\infty$ requires the partition function to approach the undeformed one. Since $\lambda^{1/2}e^{2\lambda}K_{ir}(2\lambda)$ and $\lambda^{1/2}e^{-2\lambda}I_{\pm ir}(2\lambda)$ approach constants while the other combinations grow or decay exponentially, this selects
\begin{align}\label{branch_selection}
f_r(\lambda)\propto K_{ir}(2\lambda)\quad(\epsilon=+1)\ ,\qquad\quad
f_r(\lambda)\propto I_{ir}(2\lambda)\quad(\epsilon=-1)\ .
\end{align}
On the complex-energy branch, the CFT limit leaves an admixture of $K_{ir}(2\lambda)$, suppressed by $e^{-4\lambda}$, undetermined. As we see below, it is the negative-frequency mode and is excluded by \eqref{positive_frequency}.\footnote{\label{fn:NP}The CFT limit alone would allow more than one nonperturbative completion, cf.~\cite{Aharony:2018bad,Gu:2025tpy,Gu:2026rck}. The positive-frequency mode maps to $I_{ir}(2\lambda)$, whose imaginary part, ${\rm Im}\,I_{ir}=-\pi^{-1}\sinh(\pi r)K_{ir}$, is invisible in the $1/\lambda$ expansion; it is suppressed by $e^{-4\lambda}$ for $r\ll2\lambda$ and by $e^{-2\lambda}$ for $r\gtrsim2\lambda$ (Section~\ref{sec:normalizable}). This is reminiscent of the nonperturbative sectors found for this sign of the coupling via resurgence in \cite{Gu:2025tpy}, and suggests that the choice of positive frequency fixes the corresponding resummation ambiguity. We leave a detailed comparison to future work.}

The solutions \eqref{branch_selection} admit integral representations matching the positive-frequency modes,
\begin{align}\label{Bessel_integral}
K_{ir}(2\lambda)=\frac12\int_{-\infty}^{\infty}dT\,e^{-2\lambda\cosh T}e^{-irT}\ ,\qquad\quad
I_{ir}(2\lambda)=\frac{1}{2\pi i}\int_{\mathcal C}dT\,e^{2\lambda\cosh T}e^{-irT}\ ,
\end{align}
where $\mathcal C$ is the Schl\"afli contour of Figure~\ref{fig:contour}. Choosing the normalization such that the $\lambda\to\infty$ limit reproduces the undeformed CFT partition function, we obtain the integral transform,
\begin{align}\label{forward_transform}
Z_{T\bar{T}}(\lambda,m_a)
=
\int_{\Gamma_\epsilon}dT\, K_\epsilon(\lambda, T)\,\Psi(T,m_a)\ ,
\qquad\quad
\Gamma_+=\mathbb{R}\ ,\quad \Gamma_-=\mathcal{C}\ ,
\end{align}
with the kernels
\begin{align}\label{kernels}
K_+(\lambda,T)=\frac{1}{\sqrt{\pi}}\,\lambda^{1/2}e^{-2\lambda(\cosh T-1)}\ ,\qquad\quad
K_-(\lambda,T)=\frac{1}{i\sqrt{\pi}}\,\lambda^{1/2}e^{2\lambda(\cosh T-1)}\ .
\end{align}
The mode \eqref{positive_frequency} maps to $(2/\sqrt\pi)\lambda^{1/2}e^{2\lambda}K_{ir}(2\lambda)$ on the Hagedorn branch and to $2\sqrt\pi\lambda^{1/2}e^{-2\lambda}I_{ir}(2\lambda)$ on the complex-energy branch, while $e^{+irT}$ would map to $I_{-ir}(2\lambda)=I_{ir}(2\lambda)+(2i/\pi)\sinh(\pi r)K_{ir}(2\lambda)$ on the latter.

\begin{figure}[!h]
\centering
\begin{tikzpicture}[scale=1.0,>=stealth]
\draw[->,gray] (-3.2,0) -- (5.4,0) node[right]{\small ${\rm Re}\,T$};
\draw[->,gray] (0,-2.7) -- (0,2.8) node[above]{\small ${\rm Im}\,T$};
\draw[dashed,gray] (-3.2,1.8) -- (5.2,1.8);
\draw[dashed,gray] (-3.2,-1.8) -- (5.2,-1.8);
\node[left] at (-0.05,2.08) {\small $i\pi$};
\node[left] at (-0.05,-2.08) {\small $-i\pi$};
\draw[very thick,red!70!black,dashed] (-3.0,0.06) -- (5.0,0.06);
\draw[very thick,red!70!black,->] (3.6,0.06) -- (4.2,0.06);
\node[red!70!black] at (4.3,0.38) {\small $\Gamma_+=\mathbb{R}$};
\draw[very thick,blue!70!black] (5.0,-1.8) -- (0,-1.8) -- (0,1.8) -- (5.0,1.8);
\draw[very thick,blue!70!black,->] (3.2,-1.8) -- (2.6,-1.8);
\draw[very thick,blue!70!black,->] (0,-1.15) -- (0,-0.75);
\draw[very thick,blue!70!black,->] (0,0.75) -- (0,1.15);
\draw[very thick,blue!70!black,->] (2.6,1.8) -- (3.2,1.8);
\node[blue!70!black] at (3.6,2.15) {\small $\Gamma_-=\mathcal{C}$};
\fill (0,0) circle (2pt);
\node[below right] at (0.05,-0.02) {\small $T=0$};
\end{tikzpicture}
\caption{Integration contours in the complex WdW-time plane: the real axis $\Gamma_+=\mathbb{R}$ (dashed) on the Hagedorn branch, and the Schl\"afli contour $\Gamma_-=\mathcal C$ (solid) on the complex-energy branch. The vertical segment $T=i\theta$ of $\mathcal C$ passes through the CFT point $T=0$; its horizontal legs lie at ${\rm Im}\,T=\pm\pi$. Since the integrand is analytic, each contour can be deformed without changing $Z_{T\bar T}$, as long as its ends stay in the same valleys, the asymptotic regions in which the kernel decays: ${\rm Re}\,T\to\pm\infty$ with $|{\rm Im}\,T|<\pi/2$ for $\Gamma_+$, and ${\rm Re}\,T\to+\infty$ with $\pi/2<\pm{\rm Im}\,T<3\pi/2$ for $\mathcal C$.}
\label{fig:contour}
\end{figure}

A few remarks are in order. First, the Hagedorn kernel is even in $T$, so only the even part of the wavefunction contributes, and for an even wavefunction \eqref{forward_transform} reduces to the half-line form
\begin{align}\label{half_line}
Z_{T\bar{T}}(\lambda,m_a)
=
\frac{2}{\sqrt{\pi}}\,
\lambda^{1/2}
\int_0^\infty
dT\,
e^{2\lambda(1-\cosh T)}
\Psi(T,m_a)\ .
\end{align}
For the positive-frequency mode, the half-line form fails to solve the flow equation because of a boundary term at $T=0$, and the whole real line is required. On this branch the transform does not distinguish $e^{\mp irT}$; it is one-to-one on positive-frequency states but not on all solutions of the WdW equation. On the complex-energy branch, by contrast, the two frequencies map to distinct Bessel functions.

Second, the CFT limit follows directly from the large-$\lambda$ behavior of the kernels. On the Hagedorn branch, near the peak at $T=0$, where $\cosh T-1=\frac{T^{2}}{2}+{\cal O}(T^{4})$,
\begin{equation}\label{delta}
    K_+(\lambda,T)
    \simeq
    \sqrt{\frac{\lambda}{\pi}}\,
    e^{-\lambda T^{2}+{\cal O}(\lambda T^{4})}
    \quad\xrightarrow{\lambda\to\infty}
    \quad
    \delta(T)\ .
\end{equation}
On the complex-energy branch, the same Gaussian appears along the vertical segment $T=i\theta$ of $\mathcal C$, where $K_-\,dT\simeq\sqrt{\lambda/\pi}\,e^{-\lambda\theta^2}d\theta$, while the horizontal legs are suppressed by $e^{-4\lambda}$. On both branches, the integral transform therefore localizes at $T=0$, corresponding to the asymptotic AdS$_3$ boundary, and the $T\bar T$-deformed partition function reduces to the boundary value of the WdW wavefunction, thereby recovering the standard AdS/CFT relation.

Third, since the transform acts on $T$ only, it applies to both parts of \eqref{two_parts}. For $F_{L^2}$ it holds mode by mode. For $F_{\rm light}$, the deformed partition function follows from the diffusion kernel \eqref{diffusionRG} applied to the light seeds, and the bulk wavefunction is defined by the inverse transform of Section~\ref{sec:Inverse}, continued analytically where necessary.


\subsection{Interpretation of the Kernel}
\label{sec:Kernel}

\begin{figure}[!h]
\vspace{.2cm}
\centering \includegraphics[width=\textwidth]{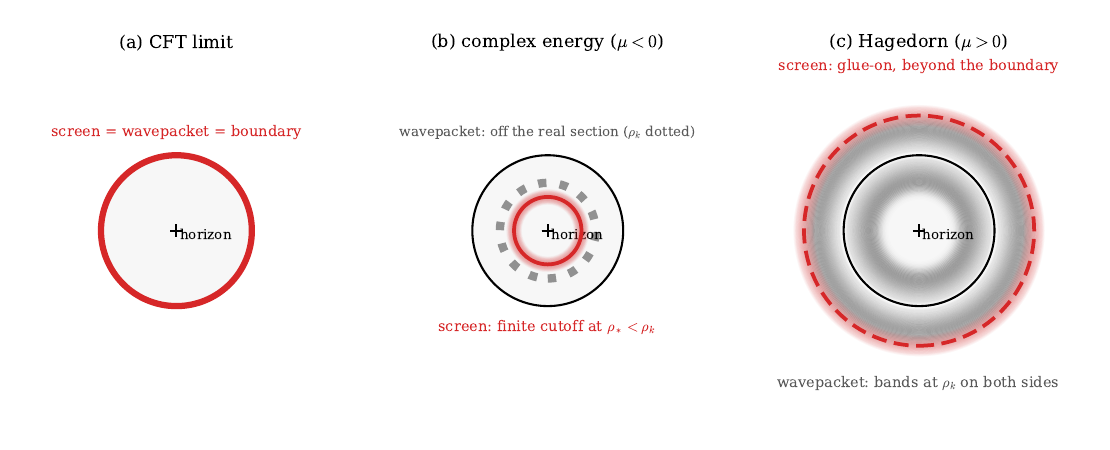}
\caption{Wavepacket and screen in proper distance from the horizon (display radius $\tanh(\rho/2)$; the glue-on side, $\rho=x+i\pi/2$, is drawn outside the boundary circle). Grey: the wavepacket, i.e.\ the weight of the kernel per unit proper distance, peaked at $\rho_k$ \eqref{radial_kernel_asym}. Red: the screen of the semiclassical state, i.e.\ the support of the integrand $K_\epsilon\Psi$. (a) In the CFT limit, both lie at the asymptotic boundary. (b) On the complex-energy branch, the wavepacket lies along the imaginary $T$ axis, i.e.\ the vertical segment of $\mathcal C$, and meets the real section only at the boundary, $T=0$ (Figure~\ref{fig:wavepacket}), so its depth $\rho_k$ on the complex section $\rho=x\mp i\pi/4$ is indicated (dotted); its weight along the two legs of $\mathcal C$ is suppressed by $e^{-4\lambda}$. The screen is not sharp: it is centered on the finite cutoff surface at $\rho_\ast<\rho_k$, with a finite width of order $(c\,m_2)^{-1/2}$ (Section~\ref{sec:semiclassical}), which vanishes only as $c\to\infty$. (c) On the Hagedorn branch, the wavepacket straddles the boundary, and the screen, again of finite width, is centered on the glue-on surface beyond it. Here $\lambda=50$, with $a=40$ in (b) and $a=20$ in (c).}
\label{fig:holographic_screen}
\end{figure}

Before examining the kernel in detail, it is useful to summarize the physical picture that emerges from the integral transform, illustrated in Figure~\ref{fig:holographic_screen}. As we shall see, the kernel implements a change of representation from the York time to its conjugate, the spatial volume. A $T\bar T$-deformed partition function at fixed coupling is therefore the amplitude of the bulk state in a fixed-volume wavepacket, which is state-independent and of finite width; in the CFT limit, it localizes at the asymptotic boundary, as in the conventional AdS/CFT correspondence. The holographic screen of a given state is instead the support of the integrand of the transform. For the semiclassical part of the bulk state, it separates from the wavepacket at large $c$ and sharpens, with a width of order $c^{-1/2}$ at fixed cutoff radius, to the finite cutoff surface of \cite{McGough:2016lol,Kraus:2018xrn} on the complex-energy branch, and to a glue-on surface beyond the asymptotic boundary \cite{Apolo:2023vnm} on the Hagedorn branch.

\subsubsection{Euclidean Rindler AdS$_3$}
\label{sec:RindlerAdS}

We first elaborate on the geometric meaning of the WdW time $T$. Setting the cosmological constant to $\Lambda=-1/\ell^2$, the classical AdS$_3$ geometry with a torus boundary in the CMC (York) gauge is given by
\begin{equation}
\begin{aligned}\label{torusU_metric1}
ds^2 = \ell^2\left(d\rho^2 +{i\over 4}(\bar{r}_1r_2-r_1\bar{r}_2)\sinh(2\rho)m_2^{-1}|dx+m dy|^2\right)\ ,
\end{aligned}
\end{equation}
where $x\sim x+1$ and $y\sim y+1$, and the torus modulus evolves along the radial direction according to
\begin{align}
m = {r_1e^{\rho}+\bar{r}_1e^{-\rho}\over r_2e^{\rho}+\bar{r}_2e^{-\rho}}\ .
\end{align}
The two complex parameters $r_1$ and $r_2$ parametrize the phase space associated with the torus modulus, while the expression above describes its classical radial evolution in AdS$_3$.
Introducing the new coordinates
\begin{align}
\alpha={1\over 2}\left((r_2+\bar{r}_2)x +(r_1+\bar{r}_1)y\right)\ ,\qquad
\beta = -{i\over 2}\left((r_2-\bar{r}_2)x +(r_1-\bar{r}_1)y\right)\ ,
\end{align}
the metric takes the form
\begin{align}\label{RindlerAdS}
ds^2 =\ell^2\left( d\rho^2 +\cosh^2\rho\, d\alpha^2+\sinh^2\rho\, d\beta^2\right)\ ,
\end{align}
which is recognized as Euclidean Rindler AdS$_3$.

The trace of the extrinsic curvature of a constant-$\rho$ hypersurface is
\begin{align}
-\tau_E\equiv K=-{2\over\ell}\coth(2\rho)\ ,
\end{align}
which implies that the WdW time $T$, defined in \eqref{define_T}, is related to the radial coordinate by
\begin{align}\label{T_rho}
T=\ln\coth\rho\ ,\qquad\mbox{or equivalently}\qquad \cosh T=\frac{\ell\tau_E}{2}\ .
\end{align}
Thus, $T$ increases monotonically from $0$ at the asymptotic AdS$_3$ boundary ($\rho\to\infty$) to $+\infty$ at the Rindler horizon ($\rho\to0$). Negative $T$ corresponds to $\rho\to\rho+i\pi/2$, which flips the sign of $e^{2\rho}$ and lies beyond the boundary.

In terms of York time, the real axis of Figure~\ref{fig:contour} covers the real slices $\tau_E\ge2/\ell$. The vertical segment $T=i\theta$ covers the classically forbidden window $|\tau_E|<2/\ell$, where $\tau_E^2+4\Lambda<0$ and, by \eqref{H_solved_E}, no real flat torus exists. The horizontal legs cover $\tau_E\le-2/\ell$, i.e.\ slices of reversed orientation. Both kernels are centered at the turning point $\tau_E=2/\ell$, the asymptotic boundary.


\subsubsection{The kernel as a change of representation}
\label{sec:representation}

In terms of the York time \eqref{T_rho}, the transform \eqref{forward_transform} reads
\begin{align}\label{Laplace_form}
Z_{T\bar{T}}(\lambda,m_a)
\propto
\lambda^{1/2}\,e^{2\epsilon\lambda}\int d\tau_E\,\frac{1}{\sqrt{\tau_E^2-4/\ell^2}}\;e^{-\epsilon\lambda\ell\tau_E}\,\Psi(\tau_E,m_a)\ ,
\end{align}
integrated over $\tau_E\ge2/\ell$ (on both sheets of the square root) for $\epsilon=+1$ (Hagedorn) and over $\tau_E\le2/\ell$ for $\epsilon=-1$ (complex energy). The exponential $e^{-\epsilon\lambda\ell\tau_E}$ implements the change of representation from the York time to its conjugate, the spatial volume, with
\begin{align}\label{V_lambda}
V=\hbar\,\ell\lambda\ ,
\end{align}
i.e.\ $V=16\pi G\ell\lambda$ upon restoring Newton's constant in \eqref{EH_ADM} and setting $\hbar=1$; the orientation of the pairing, $e^{\mp\lambda\ell\tau_E}$, is opposite on the two branches. On a Euclidean slice, this pairing is the Laplace kernel $e^{\mp V\tau_E/\hbar}$. On a Lorentzian slice, where $\tau_E$ is imaginary and the York time $\tau$ is real, it becomes the Fourier kernel $e^{iV\tau/\hbar}$, up to the time orientation: the familiar Lorentzian inner product $\langle V|\tau\rangle$ of volume and York-time eigenstates. Accordingly, a contribution weighted by $e^{-I}$ on a Euclidean section is weighted by $e^{iS}$ on a Lorentzian one, with $S$ the Lorentzian Hamilton--Jacobi function. The prefactor $e^{2\epsilon\lambda}$ is the counterterm proportional to the volume, which cancels the exponent at the boundary, and the measure is the Jacobian fixed by the operator ordering of Section~\ref{sec:CQTUAdS}, absent in \cite{Coleman:2020jte}.\footnote{\label{fn:York}Equivalently, the two representations differ by boundary terms on the slice. With the Gibbons--Hawking--York term, the on-shell variation of the Euclidean action is $\frac{1}{16\pi G}\oint(K\,\delta\sqrt g-\sqrt g\,K^{ij}\delta\hat g_{ij})$, with $\hat g_{ij}=g_{ij}/\sqrt g$, which vanishes at fixed induced metric. Subtracting half of this term trades $K\,\delta\sqrt g$ for $-\sqrt g\,\delta K$, so that the action is stationary at fixed York time and conformal class \cite{York:1986lje}; this is the action of $\Psi(\tau,m_a)$. The subtracted term, $\frac{1}{16\pi G}\oint\sqrt g\,K=\lambda\ell\tau_E$, and the counterterm, $\frac{1}{8\pi G\ell}\oint\sqrt g=2\lambda$, are the exponent and prefactor in \eqref{Laplace_form}.}

A $T\bar T$-deformed partition function at fixed coupling is therefore the bulk wavefunction in the volume representation, at fixed $V\propto m_2/|\mu|$. This is the identification underlying finite-cutoff holography, in which the induced metric is held fixed. Since the volume and the York time are conjugate, the York time, and hence the radial position, is then not sharp. This complementarity is the origin of the finite width of the holographic screen, cf.~\cite{Guica:2019nzm}.

The two branches realize this change of representation differently. On the Hagedorn branch, the transform is the projection of the Lorentzian Schr\"odinger wavefunction onto volume eigenstates, as becomes manifest on the closed universe of Section~\ref{sec:ClosedU}. On the complex-energy branch, it is a Laplace transform supported on $\tau_E\le2/\ell$, through the classically forbidden window, whose inverse is a Laplace transform from the volume to the York time as in \cite{Coleman:2020jte}, with the measure fixed by the reduced phase space (Section~\ref{sec:Inverse}).

\begin{figure}[!h]
\centering \includegraphics[width=\textwidth]{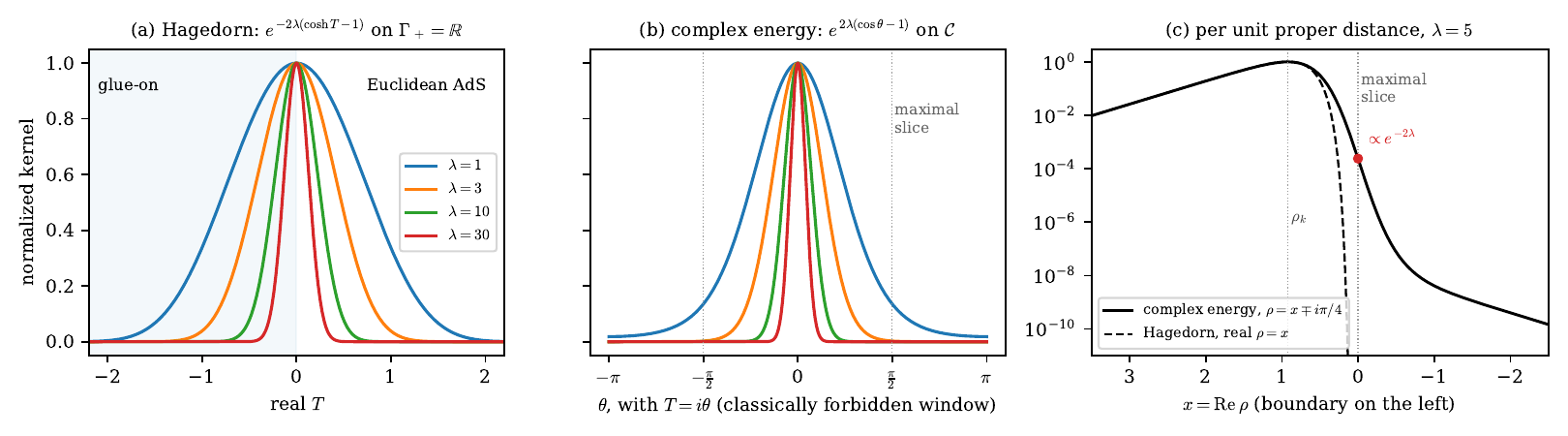}
\caption{The normalized kernels for several values of $\lambda=m_2/|\mu|$; larger $\lambda$ corresponds to weaker $T\bar T$ deformation. (a) Hagedorn branch, $e^{-2\lambda(\cosh T-1)}$ along $\Gamma_+=\mathbb R$, covering Euclidean AdS slices ($T>0$) and the glue-on region ($T<0$). (b) Complex-energy branch, $e^{2\lambda(\cos\theta-1)}$ along the vertical segment $T=i\theta$ of $\mathcal C$, through the classically forbidden window; it falls to $e^{-4\lambda}$ at the ends $\theta=\pm\pi$, where the legs of $\mathcal C$ begin. In both cases the kernel concentrates at $T=0$ and approaches a delta function in the CFT limit $\lambda\to\infty$. (c) The weight per unit proper distance for $\lambda=5$, on real $\rho$ (Hagedorn, dashed) and on the complex section $\rho=x\mp i\pi/4$ (complex energy, solid): both peak at $\rho_k$, but the complex-energy weight continues through the maximal slice $x=0$, where it is $\propto e^{-2\lambda}$.}
\label{fig:kernel}
\end{figure}

Figure~\ref{fig:kernel} illustrates the kernels \eqref{kernels} themselves, on the Hagedorn branch along $\Gamma_+$ and on the complex-energy branch along the vertical segment of $\mathcal C$, $T=i\theta$, where $\cosh T=\cos\theta$. Since $\cosh T-1=T^2/2+\cdots$, both become increasingly localized around $T=0$ as $\lambda$ increases, with the same Gaussian profile $e^{-\lambda T^2}$ and $e^{-\lambda\theta^2}$. As shown in \eqref{delta}, in the CFT limit $\lambda\rightarrow\infty$, they approach a $\delta$-function, recovering the standard AdS/CFT relation in which the torus partition function is determined by the boundary value of the bulk WdW wavefunction. For finite $\lambda$, however, the kernel is spread over a finite range of WdW times: over real slices on both sides of the boundary on the Hagedorn branch, and over the classically forbidden window on the complex-energy branch.

To obtain a more direct geometric interpretation, it is useful to
rewrite the kernel in terms of the proper radial coordinate $\rho$ of
Rindler AdS$_3$, using \eqref{T_rho}. From the half-line form \eqref{half_line}, this introduces the radial kernel
\begin{align}\label{radial_kernel}
\widetilde{K}(\lambda,\rho)
=
\frac{2}{\sqrt\pi}\lambda^{1/2}e^{2\lambda(1-\cosh T)}
\left|\frac{dT}{d\rho}\right|
=\frac{4}{\sqrt\pi}\,\frac{\lambda^{1/2}}{\sinh2\rho}\,e^{2\lambda(1-\coth2\rho)}\ ,
\end{align}
the weight per unit $\rho$, i.e.\ per unit proper distance $\ell\rho$ in units of $\ell$. The profile is
localized around a finite radial position, with its peak located at
\begin{align}
\rho_k=\frac14\sinh^{-1}(4\lambda)\ .
\end{align}
As $\lambda\rightarrow\infty$, the peak moves
monotonically toward the asymptotic AdS$_3$ boundary, recovering the CFT
limit. Its shape, however, does not change: for $\lambda\gg1$,
\begin{align}\label{radial_kernel_asym}
\widetilde{K}(\lambda,\rho)\simeq\sqrt{\frac{8}{\pi}}\,\exp\left[-2(\rho-\rho_k)-\frac12e^{-4(\rho-\rho_k)}\right]\ ,\qquad \rho_k\simeq\frac14\ln(8\lambda)\ ,
\end{align}
a fixed profile of width of order one in $\rho$, i.e.\ of order $\ell$ in proper distance, translated with the coupling. The narrowing of the kernel in $T$ as $\lambda^{-1/2}$ is thus exactly compensated by the exponential relation between $T$ and the proper distance near the boundary: in proper distance, the radial resolution of the change of representation is of order the AdS radius $\ell$, independently of the deformation.

As a function of $T$, the kernel is the York-time wavefunction of the fixed-volume state: a bulk wavepacket centered at the asymptotic boundary $T=0$, of width $\delta T_K=(2\lambda)^{-1/2}$, and $Z_{T\bar T}$ is the amplitude of $\Psi$ in it.\footnote{More precisely, of its positive-frequency part, the only part that pairs with $\Psi$.} In Euclidean York time, it is localized at the boundary $\tau_E=2/\ell$ on both branches, but from opposite sides (Figure~\ref{fig:kernel}): over the real slices $\tau_E>2/\ell$ on the Hagedorn branch, and inside the classically forbidden window $\tau_E<2/\ell$, the vertical segment of $\mathcal C$, on the complex-energy branch. On the Hagedorn branch, its weight \eqref{radial_kernel} per unit proper distance peaks at $\rho_k$ and decays toward the boundary, giving a band of width of order $\ell$ at depth $\rho_k$. On the complex-energy branch, the window consists of the complex slices $\rho=x\mp i\pi/4$ of \eqref{complex_Rindler}, along which $\ell x$ is a real proper distance, and the weight per unit $x$, $\propto{\rm sech}(2x)\,e^{-2\lambda(1-\tanh2x)}$, is the same band at depth $\rho_k$; instead of ending at a horizon, it falls to $e^{-2\lambda}$ at the maximal slice, $x=0$ (Figure~\ref{fig:holographic_screen}). This wavepacket is the probe defined by the deformation and is state-independent. The holographic screen of a given state is instead the support of the integrand $K_\epsilon\Psi$ of the transform \eqref{forward_transform}, determined by its saddle points, to which we turn in Sections~\ref{sec:semiclassical} and \ref{sec:normalizable}.

\begin{figure}[!h]
\vspace{0.3cm}
\centering \includegraphics[height=2.5in]{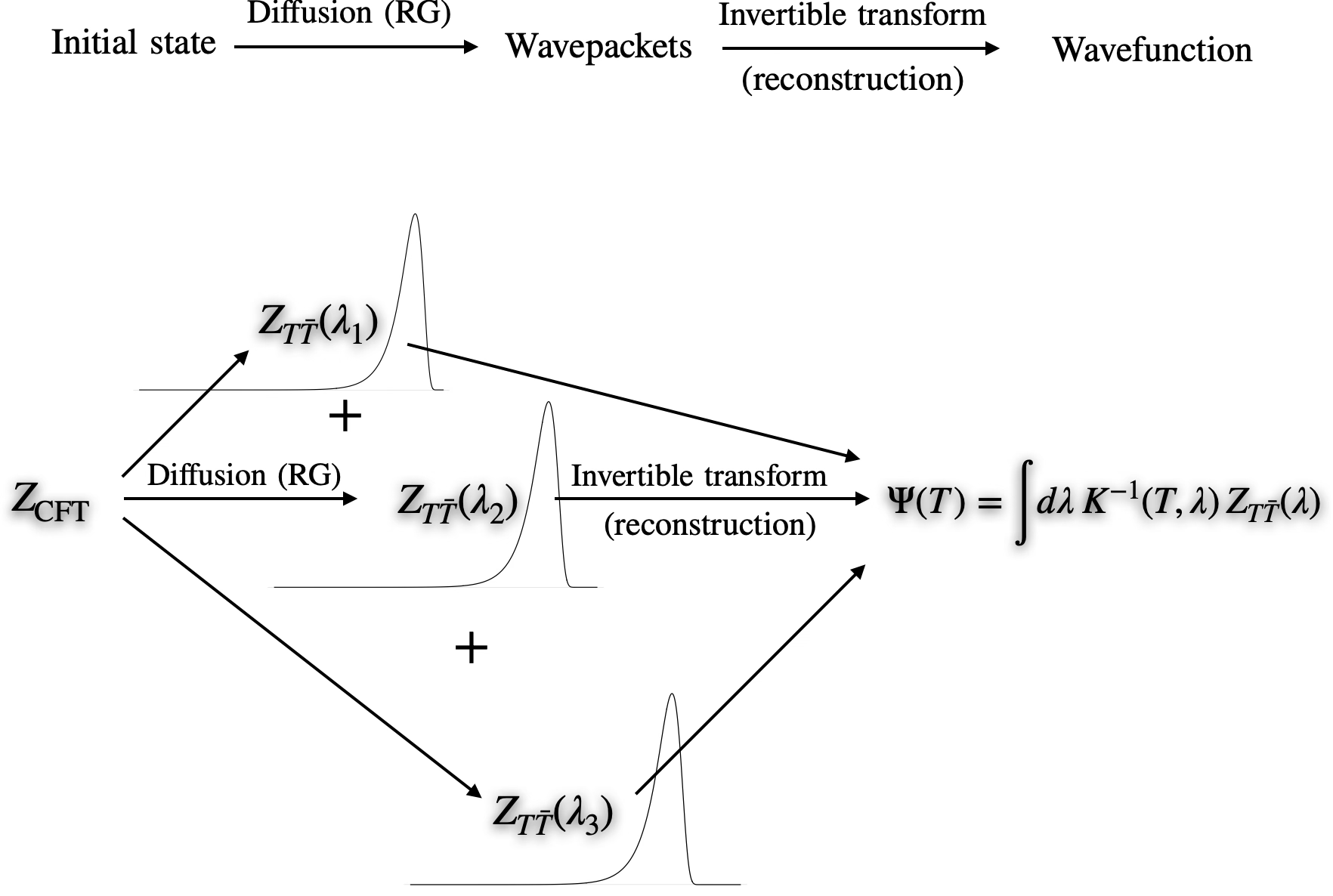}
\caption{Two-step holographic correspondence between the boundary theory and the bulk wavefunction. The diffusion (RG) transform maps the CFT partition function to the family of $T\bar{T}$-deformed partition functions $Z_{T\bar T}(\lambda)$, each the amplitude of the bulk state in a fixed-volume wavepacket. The invertible kernel transform then reconstructs the WdW wavefunction from this family; its explicit form is derived in Sec.~\ref{sec:Inverse}. The curves show the wavepackets on the Hagedorn branch for $\lambda_1<\lambda_2<\lambda_3$, against proper distance with the asymptotic boundary on the left: a common profile of width of order $\ell$, centered at $\rho_k$, which moves toward the boundary as $\lambda$ grows. On the complex-energy branch, the same profile lies on the complex slices $\rho=x\mp i\pi/4$ of the classically forbidden window.}
\label{fig:scheme}
\end{figure}

The kernel transform developed in this work suggests a new perspective on holographic bulk reconstruction, summarized schematically in Fig.~\ref{fig:scheme}. The construction naturally separates into two conceptually distinct steps. The first, relating the CFT to the $T\bar{T}$-deformed partition function, is governed by the diffusion kernel of Refs.~\cite{Dubovsky:2018bmo,Hashimoto:2019wct} and admits an RG-like interpretation, reorganizing the boundary data into a representation adapted to a bulk scale:\footnote{The integral \eqref{diffusionRG} converges on the Hagedorn branch for $0<\mu<\mu_H$, where $\mu_H$ is the Hagedorn point. Beyond it, and on the complex-energy branch, it is defined by analytic continuation in $\mu$.}
\begin{equation}\label{diffusionRG}
Z_{T\bar{T}}(\lambda,m_a)
=
\frac{\lambda}{\pi}
\int_{\mathrm{UHP}}
\frac{d^2\zeta}{\zeta_2^2}
\exp\!\left(
-\lambda
\frac{|\zeta-m|^2}{\zeta_2m_2}
\right)
Z_{\mathrm{CFT}}(\zeta_a).
\end{equation}
The second step, namely the kernel transform relating the $T\bar{T}$-deformed partition function to the bulk WdW wavefunction, is qualitatively different. Since the transform is explicitly invertible, it should not be viewed as a further RG evolution or coarse-graining, but rather as an information-preserving change of representation that identifies the boundary deformation scale with the volume, conjugate to the York time, which in Euclidean signature sets the radial position. Together, these two transforms establish a holographic dictionary between the boundary theory and the bulk wavefunction.

\subsubsection{Semiclassical holographic screens: finite cutoff and glue-on}
\label{sec:semiclassical}

We now show that, for the light part $F_{\rm light}$ at large $c$, the screen is sharply localized, but on opposite sides of the asymptotic boundary on the two branches: at the finite radial cutoff of \cite{McGough:2016lol,Kraus:2018xrn} on the complex-energy branch, and on the glue-on surface of \cite{Apolo:2023vnm} on the Hagedorn branch.

At large $c$, $F_{\rm light}$ is dominated throughout the fundamental domain by the vacuum, $E=-c/12$ and $P=0$. By \eqref{deformed_spectrum}, its deformed contribution on the complex-energy branch is
\begin{align}\label{Zvac}
Z_{\rm vac}(\lambda,m_a)=\exp\left[2\lambda\left(\sqrt{1+\frac{a}{\lambda}}-1\right)\right]\ ,
\qquad\quad
a\equiv\frac{\pi c\,m_2}{6}\ .
\end{align}
The vacuum stays real on this branch; the complex energies arise from states with $E\gtrsim1/|\mu|$.

Both $Z_{\rm vac}$ and the corresponding bulk wavefunction are determined by a single classical geometry: the smooth solution \eqref{RindlerAdS} in which the $\beta$-cycle, of period $2\pi$, shrinks at $\rho=0$. Since the two cycles scale as $\cosh\rho$ and $\sinh\rho$, its constant-$\rho$ torus has modulus $m_2$ for $\alpha\sim\alpha+2\pi m_2\tanh\rho$, and its volume is $4\pi^2\ell^2m_2\sinh^2\rho$. With the two boundary terms of footnote~\ref{fn:York}, the on-shell actions are
\begin{align}\label{two_actions}
-I_\tau=a\tanh\rho\ ,\qquad -I_V=a\left(1-e^{-2\rho}\right)\ .
\end{align}
The second is $\log Z_{\rm vac}$ at
\begin{align}\label{lambda_saddle}
\lambda=a\sinh^2\rho\ ,
\end{align}
i.e.\ at the volume of the torus, $V=\hbar\ell\lambda$, as required by \eqref{V_lambda}. The first is the classical York-time wavefunction,
\begin{align}\label{Psi_light}
\Psi_{\rm light}(T,m_a)\simeq\exp\left[\frac{\pi c\,m_2}{6}\tanh\rho\right]\ .
\end{align}

\begin{figure}[!h]
\centering \includegraphics[width=\textwidth]{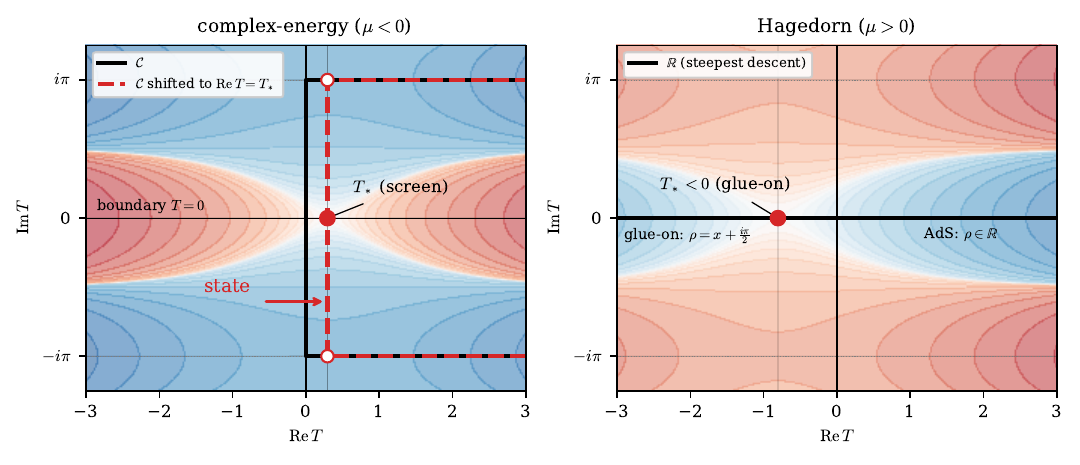}
\caption{Steepest-descent contours of the forward transform of $\Psi_{\rm light}$, for $\lambda=10$ and $a=8$, with ${\rm Re}$ of the exponent shown in color (red: larger). Left: complex-energy branch. The state shifts the vertical segment of $\mathcal C$ from the asymptotic boundary, $T=0$, to the saddle point $T_\ast>0$ (filled dot), where the contour crosses the real Euclidean section at the screen; the corners $T_\ast\pm i\pi$ (open dots) are subleading saddle points. Right: Hagedorn branch. The contour is the real axis, and the saddle point lies at $T_\ast<0$, on the glue-on side.}
\label{fig:contour_deformation}
\end{figure}

The transform reproduces this geometry as its saddle point, in either direction. Since the kernel pairs the York time with the volume, the saddle-point conditions of the transform in the two directions are the Hamilton--Jacobi relations
\begin{align}\label{HJ}
\frac{V}{\hbar}=-\partial_{\tau_E}\log\Psi\ ,\qquad \ell\tau_E=2+\partial_\lambda\log Z_{T\bar T}\ ,
\end{align}
satisfied by the actions \eqref{two_actions}. For any semiclassical state, the screen is therefore the slice of the classical geometry whose volume is fixed by the coupling, and its York time is read off from the deformed partition function. Explicitly, the inverse transform \eqref{inverse_complex} of \eqref{Zvac} is dominated by \eqref{lambda_saddle}, written as $\lambda_\ast=\frac a2(\coth T-1)$, which lies on the real $\lambda$-axis, the steepest-descent contour, and gives \eqref{Psi_light}. Conversely, the forward transform of \eqref{Psi_light} is dominated by the saddle point of $2\epsilon\lambda(1-\cosh T)+a\tanh\rho$. Since $\partial_T\tanh\rho=-\tanh\rho$, the saddle-point condition reads
\begin{align}\label{forward_saddle}
\sinh T_\ast=-\epsilon\,\frac{a\tanh\rho_\ast}{2\lambda}\ .
\end{align}
For $\epsilon=-1$, it has a real solution $T_\ast>0$ for every $\lambda>0$. With $\sinh T=1/\sinh2\rho$, it corresponds to the real radial position
\begin{align}\label{rho_cutoff}
\sinh^2\rho_\ast=\frac{\lambda}{a}=\frac{6}{\pi c|\mu|}=\frac{4G}{\pi\ell|\mu|}\ ,
\end{align}
where $c=3\ell/2G$. Along the real $T$-axis, the exponent has a minimum at $T_\ast$, with second derivative $a\sinh2\rho_\ast$, so the steepest-descent contour crosses the real axis vertically there: it is the contour $\mathcal C$ of Figure~\ref{fig:contour} with its vertical segment shifted from ${\rm Re}\,T=0$ to ${\rm Re}\,T=T_\ast$ (Figure~\ref{fig:contour_deformation}). The bulk state thus moves the point at which the contour meets the real Euclidean section from the asymptotic boundary to the screen, while the neighbouring complex geometries contribute the fluctuations.

The screen is thus the constant-$\rho_\ast$ slice of the classical geometry, and it is the finite cutoff surface of \cite{McGough:2016lol,Kraus:2018xrn}. Indeed, the deformed vacuum energy at \eqref{rho_cutoff}, $E(\mu)=-(c/6)\sinh\rho_\ast(\cosh\rho_\ast-\sinh\rho_\ast)$, satisfies
\begin{align}
2\pi m_2E(\mu)=\beta\, E_{\rm BY}(r_c)\ ,\qquad
E_{\rm BY}(r_c)=\frac{r_c}{4G\ell}\left(1-\sqrt{1+\frac{\ell^2}{r_c^2}}\right)\ ,
\end{align}
where $E_{\rm BY}$ is the Brown--York quasi-local energy \cite{Brown:1992br,Balasubramanian:1999re} of the circle of radius $r_c=\ell\sinh\rho_\ast$, and $\beta=2\pi m_2\ell\sinh\rho_\ast$ the proper length of the thermal cycle there. The deformed energy is thus the quasi-local energy at the cutoff radius $r_c^2=4G\ell/(\pi|\mu|)$, the finite-cutoff relation of \cite{McGough:2016lol,Kraus:2018xrn}. The transform therefore derives, rather than assumes, finite-cutoff holography at leading order in $1/c$.

\begin{figure}[!h]
\centering \includegraphics[width=\textwidth]{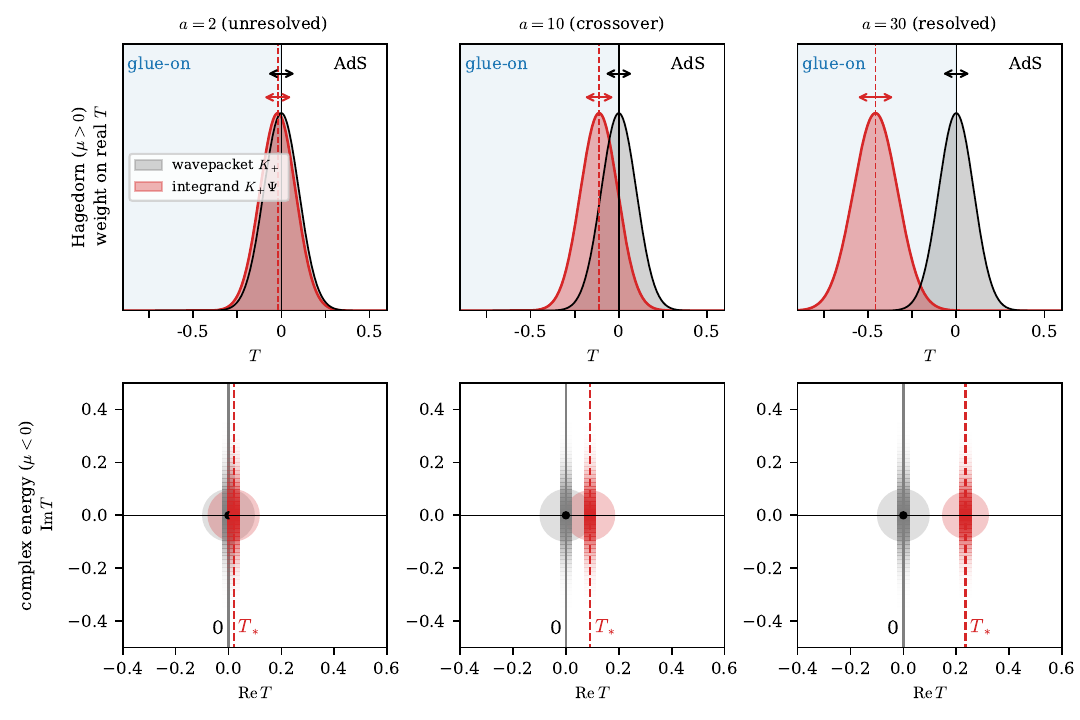}
\caption{Wavepacket and screen for the vacuum, with $\lambda=50$ and $a=2,10,30$ (left to right). Top: Hagedorn branch, weights along real $T$. The wavepacket $K_+$ (grey) is centered at the asymptotic boundary $T=0$; the integrand $K_+\Psi$ (red), whose support is the screen, moves into the glue-on region $T<0$. Bottom: complex-energy branch, in the complex $T$-plane. The Gaussian weights lie along the vertical segment of $\mathcal C$ (grey) and of its copy shifted to $T_\ast>0$ (red); the discs indicate their widths. In both cases, the screen is resolved from the wavepacket once $|T_\ast|$ exceeds the widths.}
\label{fig:wavepacket}
\end{figure}

The screen and the wavepacket are distinguished by the shift $T_\ast$ rather than by their widths. Since the exponent of the integrand is the sum of those of the wavepacket and of the state, their curvatures add, $\delta T^{-2}=2\lambda\cosh T_\ast+a\,e^{-T_\ast}$, and on the solution, where $e^{2T_\ast}=1+a/\lambda$,
\begin{align}\label{screen_width}
\delta T=\left(1+\frac{a}{\lambda}\right)^{-1/4}\delta T_K\ .
\end{align}
In WdW time, both widths are of order $(c\,m_2)^{-1/2}$ at fixed cutoff radius, since $\lambda=a\sinh^2\rho_\ast$. In proper distance, however, the wavepacket keeps its width of order $\ell$, \eqref{radial_kernel_asym}, whereas the screen, at finite $T_\ast$, has width $\ell\delta\rho=\frac{\ell}{2}\sqrt{\sinh2\rho_\ast/a}$, of order $\ell(c\,m_2)^{-1/2}$. In the semiclassical limit, $c\to\infty$ at fixed $r_c$,\footnote{This is the double-scaling limit $c\to\infty$, $\mu\to0$ at fixed $c\mu$, in which $a/\lambda=\pi c|\mu|/6$ and hence the screen stay fixed, while $\lambda$ and $a$ grow as $c$. On the Hagedorn branch, $c\mu<6/\pi$ keeps the vacuum below its branch point. At fixed $\mu$ instead, the screen moves toward the horizon on the complex-energy branch and leaves the real section past the branch point on the Hagedorn branch.} the shift $T_\ast$ stays finite while both widths shrink, so the screen separates from the wavepacket and sharpens to the cutoff surface (Figures~\ref{fig:holographic_screen} and \ref{fig:wavepacket}). In the CFT limit, $\lambda\to\infty$ at fixed $c$, the shift $T_\ast\simeq a/2\lambda$ instead falls faster than $\delta T_K$, and the screen merges with the wavepacket at the boundary. The crossover, $T_\ast\simeq\delta T_K$, i.e.\ $a^2\simeq2\lambda$, is where the $T\bar T$ correction to $\log Z_{\rm vac}\simeq a-a^2/4\lambda$ becomes of order one: the screen is resolved precisely when the deformation is semiclassically visible. In proper distance, this is $\rho_\ast<\rho_k$, the reach of the wavepacket.

On the Hagedorn branch, $\epsilon=+1$, the solution of \eqref{forward_saddle} has $T_\ast<0$. By \eqref{T_rho}, real $\rho$ gives only $T>0$, so that $\rho_\ast=x_\ast+i\pi/2$ with real $x_\ast$, for which $\coth\rho_\ast=\tanh x_\ast<1$. Here the contour is the real axis itself, which is the steepest-descent contour. With $\sinh T=-1/\sinh2x$ for $T=\ln\tanh x$, \eqref{forward_saddle} gives
\begin{align}\label{glueon}
\cosh^2x_\ast=\frac{\lambda}{a}=\frac{6}{\pi c\,\mu}\ ,
\end{align}
i.e.\ $e^{2T_\ast}=1-a/\lambda$, so that the screen is resolved under the same condition, now on the glue-on side.
The saddle point now lies beyond the asymptotic boundary, where $e^{2\rho}\to-e^{2\rho}$. The checks above continue analytically. With $r_c=\ell\sinh\rho_\ast=i\ell\cosh x_\ast$ and $\beta=2\pi m_2\ell\sinh\rho_\ast$, the deformed vacuum energy on this branch, $E(\mu)=-(c/6)\cosh x_\ast(\cosh x_\ast-\sinh x_\ast)$, again satisfies $2\pi m_2E(\mu)=\beta E_{\rm BY}(r_c)$. The volume of the continued torus, $-4\pi^2\ell^2m_2\cosh^2x_\ast$, agrees with \eqref{V_lambda} up to a sign, which reflects the reversed orientation of the pairing in \eqref{Laplace_form} and is in line with the signed areas of glue-on holography \cite{Apolo:2023ckr}. Both branches are therefore captured by the single relation
\begin{align}\label{rc_unified}
r_c^2=-\frac{4G\ell}{\pi\mu}\ ,
\end{align}
which places the screen at a finite cutoff inside the bulk for $\mu<0$ and on a glue-on surface beyond the asymptotic boundary for $\mu>0$ \cite{Apolo:2023vnm}. On the Hagedorn branch, \eqref{glueon} has a solution only for $\mu\le6/(\pi c)$. At $\mu=6/(\pi c)$, the glue-on surface reaches $x_\ast=0$ and the deformed vacuum, $\exp[2\lambda(1-\sqrt{1-a/\lambda})]$, develops a branch point; beyond it, the vacuum energy becomes complex. This is the $S$-transform of the Hagedorn point, $\mu_H=6m_2^2/(\pi c)$: the two coincide for $m_2=1$, and the regime beyond it is reached only by analytic continuation in $\mu$.

By \eqref{HJ}, the same analysis applies to the other contributions to $F_{\rm light}$. For a light primary with $P=0$ and $E=-\gamma c/12$, $\gamma=1-12\Delta/c$, the dual geometry is the conical defect in which the $\beta$-cycle has period $2\pi\sqrt\gamma$, and $a\to\gamma a$. The screen then lies at $\sinh^2\rho_\ast=6/(\pi c\gamma|\mu|)$, a depth that depends on the state, whereas its intrinsic radius, $r_c=\sqrt\gamma\,\ell\sinh\rho_\ast$, does not: \eqref{rc_unified} holds for every light state, and on the Hagedorn branch the glue-on surface exists for $\mu\le6/(\pi c\gamma)$. The coupling thus fixes the induced metric of the screen, while the state determines its embedding, i.e.\ its York time and depth, as in the finite-cutoff picture of \cite{McGough:2016lol,Kraus:2018xrn}. Likewise, each modular image of the vacuum is obtained with $a\to\pi c\,m_2/(6|cm+d|^2)$ and $\mu\to\mu/|cm+d|^2$; on the complex-energy branch, the BTZ contribution, for instance, yields $\sinh^2\rho'_\ast=6|m|^2/(\pi c|\mu|)$ in the Rindler coordinate of the $S$-transformed solution. In pure AdS$_3$ gravity, the real Euclidean saddles with torus boundary are exactly these solid tori, whose contributions form the Poincar\'e series of the vacuum \cite{Maloney:2007ud}, together with the conical-defect geometries of light primaries. All real Euclidean geometries thus reside in $F_{\rm light}$.


\subsubsection{The normalizable sector}
\label{sec:normalizable}

The normalizable part $F_{L^2}$ does not define a second screen. Since the transform is linear, the support of the bulk state is the union of the supports of its two parts, weighted by their amplitudes, and at large $c$ it is dominated by the light part, whose saddle point is the real Euclidean screen of Section~\ref{sec:semiclassical}. The part $F_{L^2}$ is subleading and, having real $r$, has no support on real Euclidean geometries (Section~\ref{sec:CQTUAdS}). Its saddle points, determined below, are complex or Lorentzian stationary points, which organize the contributions beyond the classical saddle point: quantum fluctuations, tunnelling, and the nonperturbative imaginary part of $Z_{T\bar T}$.

For the mode \eqref{positive_frequency}, the integrand of \eqref{forward_transform} is $e^{-2\epsilon\lambda(\cosh T-1)-irT}$, which is stationary at
\begin{align}\label{mode_saddle}
\sinh T_\ast=-i\epsilon\,\frac{r}{2\lambda}\ .
\end{align}
Two regimes are separated by $r=2\lambda$, the turning point of \eqref{Bessel}. On the complex-energy branch, they separate the perturbative and nonperturbative parts of the mode, $I_{ir}(2\lambda)$. Below the threshold, it is dominated by the vertical segment of $\mathcal C$, on which the integrand is real; this real part reproduces the $1/\lambda$ expansion, while the imaginary part, from the legs of $\mathcal C$, is exponentially small. Above the threshold, it comes from Lorentzian saddle points and is of order $e^{-2\lambda}$.

Below the threshold, $r<2\lambda$, one has $T_\ast=-i\epsilon\theta_\ast$ with $\sin\theta_\ast=r/2\lambda$, i.e.
\begin{align}\label{complex_rho}
\rho_\ast=\frac12\cosh^{-1}\!\left(\frac{2\lambda}{r}\right)+\frac{i\epsilon\pi}{4}\ ,
\end{align}
at the York time $\tau_{E\ast}=(2/\ell)\sqrt{1-(r/2\lambda)^2}$ inside the classically forbidden window. Setting $\rho=x-i\pi/4$ in \eqref{RindlerAdS} gives the complex geometry
\begin{align}\label{complex_Rindler}
ds^2=\ell^2\left[dx^2+\frac12\left(1-i\sinh2x\right)d\alpha^2-\frac12\left(1+i\sinh2x\right)d\beta^2\right]\ ,
\end{align}
which has an imaginary volume element and no horizon. At its throat $x=0$, i.e.\ $\tau_E=0$, it becomes real and, after $\beta\to i\beta$, is the maximal slice of the closed universe of Section~\ref{sec:ClosedU}. Like real Lorentzian metrics, it lies on the boundary of the domain of allowable complex metrics \cite{Kontsevich:2021dmb,Witten:2021nzp}. As $r$ increases to $2\lambda$, the saddle point moves from the boundary to the throat.

\begin{figure}[!htbp]
\centering \includegraphics[width=0.92\textwidth]{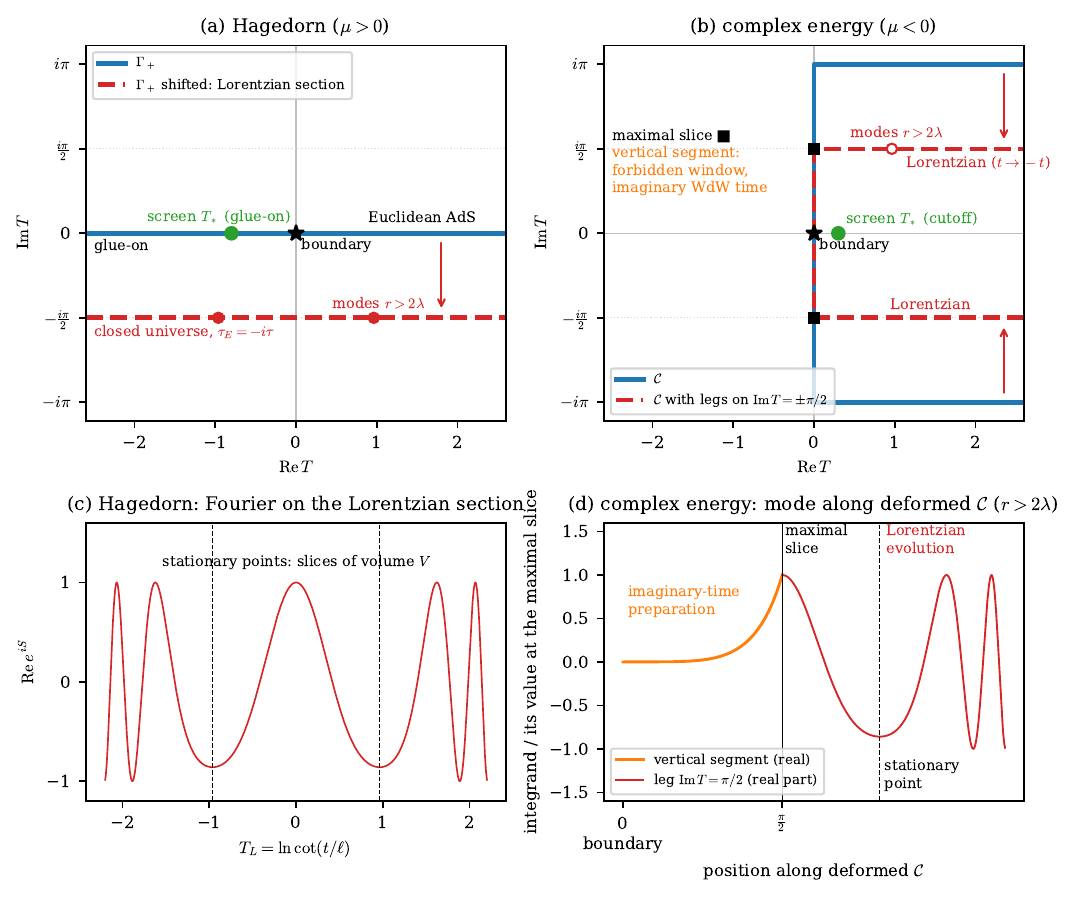}
\caption{Contours and sections of the complex $T$-plane. The real axis contains the Euclidean AdS slices ($T>0$) and the glue-on region ($T<0$), the vertical segment the classically forbidden window, and the lines ${\rm Im}\,T=\mp\pi/2$ the Lorentzian closed universe and its time reversal. Green dots: screens $T_\ast$; red dots: mode saddle points for $r>2\lambda$; squares: maximal slices. The contours of Figure~\ref{fig:contour} are moved, within their valleys, onto the edges ${\rm Im}\,T=\pm\pi/2$, where the integral is defined by the $i\epsilon$ prescription. (a) Hagedorn branch: $\Gamma_+$ is moved onto the Lorentzian section. (b) Complex-energy branch: the legs of $\mathcal C$ are moved, so that the vertical segment runs in imaginary WdW time from the boundary to the maximal slices, as in the Hartle--Hawking contour; the mode saddle points lie on the upper leg. (c) A mode with $r>2\lambda$ ($\lambda=4$, $r=12$) along the Lorentzian section of (a): it oscillates except near its stationary points, the slices of volume $V$. (d) The same mode along the upper half of the deformed $\mathcal C$ of (b), normalized at the maximal slice: real and growing along the vertical segment, the imaginary-time preparation, and oscillating along the leg, the Lorentzian evolution, with a stationary point at a slice of volume $V$.}
\label{fig:sections}
\end{figure}

Above the threshold, $r>2\lambda$, one has $T_\ast=-i\epsilon\pi/2\pm\cosh^{-1}(r/2\lambda)$. Continuing \eqref{RindlerAdS} through the horizon by $\rho=it/\ell$, together with $\beta\to i\beta$, gives
\begin{align}\label{interior_metric}
ds^2=-dt^2+\ell^2\cos^2(t/\ell)\,d\alpha^2+\ell^2\sin^2(t/\ell)\,d\beta^2\ ,
\end{align}
the region beyond the horizon of Lorentzian Rindler AdS$_3$; since both cycles are compact, it is the closed torus universe of Section~\ref{sec:ClosedU}. As $\coth(it/\ell)=-i\cot(t/\ell)$, this Lorentzian section is the line ${\rm Im}\,T=-\pi/2$, with $T_L\equiv\ln\cot(t/\ell)={\rm Re}\,T$, and its time-reversed copy is the line ${\rm Im}\,T=+\pi/2$. For $\epsilon=+1$ (Hagedorn), $\Gamma_+$ can be shifted onto the section, since the integrand is entire and decays in between. There $\tau_E=-i\tau$: the kernel becomes the Lorentzian pairing $e^{2\lambda}e^{iV\tau/\hbar}$, and the mode $e^{-irT}=e^{-\pi r/2}e^{-irT_L}$ evolves in Lorentzian time. The integrand is thus $e^{iS}$, and its saddle points are classical Lorentzian slices, at the York time $|\tau_\ast|=(2/\ell)\sqrt{(r/2\lambda)^2-1}$. For $\epsilon=-1$ (complex energy), they lie on the time-reversed copy, to which the upper leg of $\mathcal C$ can be lowered (Figure~\ref{fig:sections}). The magnitude of these contributions is $e^{-\epsilon\pi r/2}$.

On the complex-energy branch, these Lorentzian contributions carry the imaginary part of $Z_{T\bar T}$. This is the nonperturbative part noted below \eqref{branch_selection} (footnote~\ref{fn:NP}). It is the bulk counterpart of the complex spectrum. For a single state with $P=0$, \eqref{HJ} gives $\ell\tau_E=(2-u)/\sqrt{1-u}$ with $u=2\pi|\mu|E$: as $E$ approaches $1/(2\pi|\mu|)$, where its deformed energy becomes complex, its saddle point reaches a horizon, $\tau_E\to\infty$, and beyond it lies on \eqref{interior_metric}, with imaginary $\tau_E$. This is the mechanism of Cauchy slice holography \cite{Araujo-Regado:2022gvw}, which associates imaginary energies beyond the branch point with Cauchy slices behind the horizon. For the torus, however, the vacuum and its modular images never cross the branch point on this branch, since $E=-c/12<0$, so these slices are encoded only in the imaginary part, suppressed by $e^{-2\lambda}=e^{-V/8\pi G\ell}$. On the Hagedorn branch, the leading saddle itself becomes Lorentzian, but only beyond the vacuum branch point $\mu=6/(\pi c)$, i.e.\ past the Hagedorn point (Section~\ref{sec:semiclassical}). On either branch, the slices behind the horizon are thus not reached by the leading saddle within the physical range of the coupling.

In the third-quantized description of \cite{Hirano:2026vcu}, the classical screen belongs to the fixed background, while these subleading contributions are those of the universe-field modes.


\subsubsection{The holographic screen on the two branches}
\label{sec:two_screens}

The holographic interpretation of the transform thus differs sharply between the two signs of the coupling, as summarized in Table~\ref{tab:branches}. On the complex-energy branch, the screen is a finite cutoff surface inside the bulk, at $r_c^2=4G\ell/(\pi|\mu|)$, for every coupling. The normalizable sector adds the perturbative corrections, through complex saddle points below the threshold $r=2\lambda$, and the nonperturbative imaginary part of the partition function, through Lorentzian saddle points behind the horizon, mirroring the complex spectrum. On the Hagedorn branch, the screen is a glue-on surface beyond the asymptotic boundary, at $r_c^2=-4G\ell/(\pi\mu)$, up to the vacuum branch point $\mu=6/(\pi c)$, the $S$-transform of the Hagedorn point. The normalizable modes above the threshold are supported on the real Lorentzian section of the closed universe, where the transform is the fixed-volume projection of the Schr\"odinger wavefunction (Section~\ref{sec:ClosedU}), and their contributions are real.

In both cases, the kernel is a wavepacket centered at the asymptotic boundary in York time. The side of the boundary on which the screen lies is determined not by the wavepacket but by the bulk state, through the saddle points of the integrand $K_\epsilon\Psi$, and the screen is resolved from the wavepacket precisely when $a^2\gg2\lambda$, i.e.\ when the $T\bar T$ correction to $\log Z_{\rm vac}$, $a^2/4\lambda=\pi^2c^2m_2|\mu|/144$, is large (Figure~\ref{fig:wavepacket}). This is the precise sense in which the cutoff and glue-on pictures \cite{McGough:2016lol,Kraus:2018xrn,Apolo:2023vnm} are the two semiclassical limits of a single exact transform.

\begin{table}[!h]
\centering\small
\renewcommand{\arraystretch}{1.25}
\begin{tabular}{|l|c|c|}
\hline
 & \textbf{Hagedorn} ($\mu>0$) & \textbf{Complex energy} ($\mu<0$) \\
\hline
CFT limit selects & $K_{ir}(2\lambda)$ & $I_{ir}(2\lambda)$ \\
\hline
Contour & real axis & Schl\"afli contour $\mathcal C$ \\
\hline
Change of representation & fixed-volume projection & Laplace transform through \\
 & of the Lorentzian state & the classically forbidden window \\
\hline
Semiclassical screen & glue-on, $r_c^2=-4G\ell/(\pi\mu)$ & finite cutoff, $r_c^2=4G\ell/(\pi|\mu|)$ \\
 & (for $\mu\le6/(\pi c)$) & (for all $\mu$) \\
\hline
Modes with $r<2\lambda$ & complex, ${\rm Im}\,\rho_\ast=\pi/4$ & complex, ${\rm Im}\,\rho_\ast=-\pi/4$; \\
 & & perturbative part \\
\hline
Modes with $r>2\lambda$ & Lorentzian section, & time-reversed Lorentzian section, \\
 & magnitude $e^{-\pi r/2}$, real & magnitude $e^{\pi r/2}$; ${\rm Im}\,Z\sim e^{-2\lambda}$ \\
\hline
Nonperturbative ambiguity & none & fixed by positive frequency \\
\hline
\end{tabular}
\caption{The holographic interpretation of the integral transform on the two branches.}
\label{tab:branches}
\end{table}


\subsubsection{Closed Torus Universes without Asymptotic Boundaries}
\label{sec:ClosedU}

As in Section~\ref{sec:normalizable}, the Lorentzian closed AdS$_3$ torus universe is obtained from Euclidean Rindler AdS$_3$ by the double Wick rotation
\begin{align}\label{double_Wick}
\rho\to it/\ell \ ,\qquad \beta\to i\beta\ ,
\end{align}
which yields
\begin{align}\label{AdS3_torus_universe}
ds^2 =-dt^2 +\ell^2\cos^2(t/\ell)\,d\alpha^2+\ell^2\sin^2(t/\ell)\, d\beta^2\ .
\end{align}
Introducing the new coordinates
\begin{align}
\alpha={1\over 2}\left((r_2^++r_2^-)x +(r_1^++r_1^-)y\right)\ ,\qquad
\beta = {1\over 2}\left((r_2^+-r_2^-)x +(r_1^+-r_1^-)y\right)\ ,
\end{align}
the metric takes the form \cite{Carlip:2004ba}
\begin{equation}
\begin{aligned}\label{torusU_metric2}
ds^2 = -dt^2 +{\ell^2\over 4}(r_1^-r_2^+-r_1^+r_2^-)\sin(2t/\ell)\tau_2^{-1}|dx+\tau dy|^2\ ,
\end{aligned}
\end{equation}
where the torus modulus evolves in time according to
\begin{align}
\tau = {r_1^-e^{it/\ell}+r_1^+e^{-it/\ell}\over r_2^-e^{it/\ell}+r_2^+e^{-it/\ell}}\ .
\end{align}
The four real parameters $r_i^{\pm}$ $(i=1,2)$ parametrize the phase space associated with the torus modulus.
The spacetime represents a closed AdS$_3$ torus universe with cosmological
time $0<t<\pi\ell/2$. It emerges from a big bang singularity at $t=0$,
undergoes expansion until reaching a moment of maximal volume at
$t=\pi\ell/4$, and then recontracts to a big crunch singularity at
$t=\pi\ell/2$.

The trace of the extrinsic curvature of a constant-$t$ hypersurface is
\begin{align}
-\tau\equiv K=-{2\over\ell}\cot(2t/\ell)\ ,
\end{align}
which implies that the WdW time, defined in \eqref{define_T}, is related to the cosmological time by
\begin{align}\label{T_L}
T_L=\ln\cot(t/\ell)\ ,\qquad\mbox{or equivalently}\qquad \sinh T_L=\frac{\ell\tau}{2}\ .
\end{align}
Thus, $T_L$ decreases monotonically from $+\infty$ at the initial singularity ($t\to0$) to $-\infty$ at the final singularity ($t\to\pi\ell/2$), passing through $T_L=0$ at the moment of maximal expansion ($t=\pi\ell/4$).

Since \eqref{define_T} determines the WdW time only up to an additive constant, \eqref{T_L} involves a choice. The WdW equation is invariant under shifts of $T$, but the kernels \eqref{kernels} are not, so the constant determines which geometry sits at the CFT point $T=0$. It is fixed by requiring the transform to be the change of representation of Section~\ref{sec:representation}, i.e.\ the exponent of the kernel to be linear in the York time: $\cosh T_L=\sqrt{1+\ell^2\tau^2/4}$ is not, whereas $\cosh(T_L\mp i\pi/2)=\mp i\ell\tau/2$ is. The same constant follows from requiring a single analytic wavefunction in both signatures, since $\sinh^{-1}(i\ell\tau_E/2)=\cosh^{-1}(\ell\tau_E/2)\pm i\pi/2$, and it is the shift induced by the double Wick rotation, $\ln\coth(it/\ell)=\ln\cot(t/\ell)\mp i\pi/2$. The remaining sign reflects the time orientation; we take
\begin{align}\label{T_shift}
T=T_L-\frac{i\pi}{2}\ ,
\end{align}
for which the Hagedorn transform projects the positive-frequency state onto positive-volume eigenstates, as shown below.

The CFT point $T=0$ then lies at $T_L=i\pi/2$, i.e.\ $t\to-i\infty$: it is the asymptotic Euclidean boundary of the complexified universe, not a point on its real section. The maximal slice, $T_L=0$ (or $T_L=i\pi$ on the second sheet), lies at $T=\mp i\pi/2$ on the vertical segment of $\mathcal C$, where the complex geometry \eqref{complex_Rindler} meets the Lorentzian section.

On the real Lorentzian section, $\cosh T=-i\sinh T_L$, and the Hagedorn kernel becomes the Lorentzian pairing $e^{2\lambda}e^{iV\tau/\hbar}$ of Section~\ref{sec:representation}. Since $\Gamma_+$ can be shifted onto this section (Section~\ref{sec:normalizable}), the Hagedorn transform is exactly the Fourier transform of the Lorentzian Schr\"odinger wavefunction from York time to volume, i.e.\ its amplitude at fixed $V=\hbar\ell\lambda$, in the same way that a fixed-energy amplitude follows from a time-dependent wavefunction. It is not localized on a single Cauchy slice; semiclassically, it is dominated by the slices of volume $V$, as for the normalizable modes of Section~\ref{sec:normalizable}, whereas the semiclassical screen lies off this section, on the glue-on side (Figure~\ref{fig:sections}).

On the complex-energy branch, once its legs are moved onto ${\rm Im}\,T=\pm\pi/2$, the contour $\mathcal C$ has the form of the Hartle--Hawking contour \cite{Hartle:1983ai}. In the WdW time $T_L=T+i\pi/2$ of the closed universe, its vertical segment is imaginary-time evolution through the classically forbidden window, from the Euclidean boundary, $\tau_E=2/\ell$, to the turning point $\tau_E=0$, the maximal slice, while its legs are Lorentzian evolution away from the maximal slice, in the two time orientations. This is the structure of a Hartle--Hawking state prepared at its moment of time symmetry. The kernel weight falls to $e^{-2\lambda}$ at the maximal slice, and the factors $e^{\mp\pi r/2}$ of Table~\ref{tab:branches} are the imaginary-time evolution of a mode over the interval $\pi/2$ between the Euclidean and Lorentzian sections. Unlike the no-boundary state, however, the preparation starts at the asymptotic boundary rather than at a regular cap, and the intermediate geometry \eqref{complex_Rindler} is complex rather than Euclidean.

Although the Lorentzian geometry is spatially compact and has no asymptotic boundary, the maximal slice therefore retains a distinguished role, as the locus at which the complexified geometry connected to the CFT point meets the real universe. This is reminiscent of Cauchy-slice holography \cite{Araujo-Regado:2022gvw}, in which holographic data are associated with a spacelike Cauchy surface rather than an asymptotic boundary, and suggests that observables anchored on the maximal slice play a natural role in the holographic description of closed universes. Unlike the late-time cosmological correlators usually considered in de Sitter holography, these observables are associated with a distinguished finite-time Cauchy surface. Their physical interpretation deserves further investigation.



\section{Holographic Reconstruction of Bulk Wavefunctions}
\label{sec:Inverse}

The interpretation developed in the previous section suggests that a
$T\bar{T}$-deformed partition function should be viewed not as the WdW
wavefunction evaluated on a single bulk slice, but as a bulk wavepacket of
finite radial (or York-time) extent. The crucial feature of the kernel
transform is that it is invertible. The complete bulk WdW wavefunction is
therefore encoded in the entire family of $T\bar{T}$-deformed partition
functions, each representing a wavepacket whose radial reach is set by the
coupling. In this section, we derive the explicit inverse transform, thereby
reconstructing the bulk wavefunction from these wavepackets and completing
the holographic dictionary.

On the complex-energy branch, the identity $\int_0^\infty dx\,e^{-x\cosh T}I_{ir}(x)=e^{-irT}/\sinh T$ recovers the positive-frequency modes directly, and the bulk wavefunction is reconstructed as
\begin{align}\label{inverse_complex}
\Psi(T,m_a)
=
\frac{\sinh T}{\sqrt\pi}
\int_0^\infty
d\lambda\,
\lambda^{-1/2}
e^{2\lambda(1-\cosh T)}
Z_{T\bar{T}}(\lambda,m_a)\qquad(\epsilon=-1)\ ,
\end{align}
for $T>0$, and by analytic continuation elsewhere. This is a Laplace transform from the volume to the York time, as in \cite{Coleman:2020jte}, with the measure fixed by the reduced phase space. For $F_{\rm light}$, it defines $\Psi_{\rm light}$, with the integral continued analytically where it does not converge.

On the Hagedorn branch, we use the Laplace transform identity
for the modified Bessel function,
\begin{align}
\frac{2}{\pi}\sinh T
\int_0^{\infty} d\lambda\,
e^{-2\lambda\cosh T}
K_{ir}(2\lambda)
=
\frac{\sin(rT)}{\sinh(\pi r)}.
\end{align}
Defining
\begin{align}\label{G_def}
G(T,m_a)
\equiv
\frac{\sinh T}{\sqrt\pi}
\int_0^\infty
d\lambda\,
\lambda^{-1/2}
e^{-2\lambda(1+\cosh T)}
Z_{T\bar{T}}(\lambda,m_a)\qquad(\epsilon=+1)\ ,
\end{align}
each positive-frequency component $a_re^{-irT}$ of the wavefunction yields $G_r=a_r\sin(rT)/\sinh(\pi r)$. The bulk wavefunction is therefore reconstructed as
\begin{align}\label{reconstruction}
\Psi(T,m_a)
=
\sinh(\pi\hat r)\left(\hat r^{-1}\partial_T-i\right)G(T,m_a)\ ,\qquad
\hat r\equiv\sqrt{-\left(\Delta_{\rm Maass}+\tfrac14\right)}\ ,
\end{align}
where $\hat r$ acts on the moduli. Equivalently, $\sinh(\pi\hat r)\,\hat r^{-1}\partial_TG=\sin(\pi\partial_T)G$, with the shift operator acting on the analytic continuation of $G$ in $T$. It cannot be taken inside the $\lambda$-integral, where the shifted integrand $e^{2\lambda(\cosh T-1)}$ grows and, acting on the kernel alone, the shift would give zero.
Equations \eqref{inverse_complex} and \eqref{reconstruction} establish that the kernel transform is
one-to-one: the family of $T\bar{T}$-deformed partition functions contains
the same information as the bulk WdW wavefunction. While a partition function
at fixed deformation parameter captures only a finite-resolution component of
the bulk state, the complete one-parameter family provides an equivalent
representation from which the exact wavefunction is recovered by the inverse
transform.

This reconstruction should be distinguished from the diffusion transform
\eqref{diffusionRG}, which maps the undeformed CFT partition function to the
family of $T\bar{T}$-deformed partition functions. The diffusion transform has
a natural RG interpretation, reorganizing the boundary description according
to the deformation scale. By contrast, the kernel transform derived here is an
invertible change of representation rather than a further coarse-graining.
Taken together, these two transforms naturally separate the holographic
correspondence into two stages: the first reorganizes the boundary data into a
scale-dependent representation, while the second reconstructs the bulk WdW
wavefunction from that representation.

From this viewpoint, the one-parameter family of $T\bar{T}$-deformed partition
functions may be regarded as a tomographic encoding of the bulk quantum state.
Different values of the deformation parameter provide complementary
finite-resolution probes of the same state, and only the complete family
contains sufficient information to recover the exact wavefunction. Unlike
HKLL reconstruction or entanglement wedge reconstruction, which focus on local
bulk operators or operator algebras, the present construction reconstructs the
WdW wavefunction itself. The object of reconstruction is therefore the bulk
quantum state rather than individual observables.

\section{de Sitter Torus Universes}
\label{sec:dS}

Much of the preceding analysis extends naturally from asymptotically AdS$_3$ spacetimes to asymptotically dS$_3$ universes. In fact, the canonical quantization developed in Section~\ref{sec:CQTUAdS} and the derivation of the kernel transform in the first part of Section~\ref{sec:MapWFUandTTbar} were carried out without fixing the sign of the cosmological constant. The sign of $\Lambda$ enters only through the relation between the WdW time $T$ and the York time $\tau$ (or $\tau_E$) in \eqref{define_T}. Consequently, the same kernel transform defines a holographic dictionary between the dS$_3$ bulk WdW wavefunction and the $T\bar{T}$-deformed torus partition function. A correspondence between the dS$_3$ wavefunction and the $T\bar{T}$-deformed torus partition function was previously proposed in~\cite{Godet:2024ich}, although the holographic dictionary developed here is different. See also~\cite{Chakravarty:2025sbg} for related work on bulk wavefunctions. More broadly, our work contributes to the growing body of work relating $T\bar{T}$ deformations and de Sitter holography, beginning with~\cite{Gorbenko:2018oov}.

One important question is the geometric realization of the CFT limit,
$T=0$, or equivalently, the identification of the corresponding
``holographic screen,'' in the dS$_3$ picture. In the Euclidean
AdS$_3$ interpretation, the CFT limit is associated with the asymptotic
boundary, whereas in the Lorentzian AdS$_3$ torus universe it lies off the real
section, the maximal slice instead marking the locus at which the complexified
geometry meets the real universe. We now determine the corresponding
loci in the dS$_3$ universe.

\subsection{Lorentzian universes}
\label{sec:dS_Lorentzian}

The dS$_3$ torus universe is obtained from the AdS$_3$ closed torus
universe \eqref{AdS3_torus_universe} by the analytic continuations
$\ell\to i\ell$ and $\alpha\to i\alpha$, yielding
\begin{align}\label{dS3torus}
ds^2
=
-dt^2
+\ell^2\cosh^2(t/\ell)\,d\alpha^2
+\ell^2\sinh^2(t/\ell)\,d\beta^2\ .
\end{align}
The cosmological time ranges over $0\le t<\infty$
where $t=0$ corresponds to the initial cosmological singularity at which the
$\beta$-cycle shrinks to zero size. As $t\to\infty$, the geometry approaches
future infinity of the dS$_3$ torus universe.

The trace of the extrinsic curvature of a constant-$t$ hypersurface is
\begin{align}
\tau\equiv-K
=
\frac{2}{\ell}\coth(2t/\ell)\ .
\end{align}
Substituting this into the definition of the WdW time \eqref{define_T}
with $\Lambda=1/\ell^2$, we obtain
\begin{align}
T
=
\ln\coth(t/\ell)\ .
\end{align}
Thus, $T$ decreases monotonically from $+\infty$ at the initial
singularity ($t\to0$) to $0$ at future infinity ($t\to\infty$). The
CFT limit is therefore realized at future infinity of the dS$_3$ torus
universe. In the language introduced in the previous section, future
infinity thus plays the role of the corresponding ``holographic
screen.'' This is the natural analogue of the asymptotic boundary in
Euclidean AdS$_3$ and the maximal Cauchy slice in the AdS$_3$ closed
torus universe. 

This identification agrees with the conventional
Lorentzian picture of the dS/CFT correspondence \cite{Strominger:2001pn}.
Since $\cosh T=\ell\tau/2$, the dS$_3$ torus universe is the counterpart of
Euclidean AdS$_3$: its real slices cover $\tau\ge2/\ell$, and the window
$|\tau|<2/\ell$ is classically forbidden. On this section, the Hagedorn kernel
is the real exponential $e^{-\lambda\ell\tau}$, so that real $\lambda$
corresponds to an imaginary volume, $V=-i\hbar\ell\lambda$, as expected from
the continuation $\ell\to i\ell$ relating the AdS$_3$ and dS$_3$ universes.
As we shall see next, the Euclidean representation obtained by a double Wick
rotation instead contains the maximal-volume torus, as the closed AdS$_3$
universe does.

Beyond the CFT limit, the analogue of the finite cutoff is less clear, since the continuation $\ell\to i\ell$ makes the central charge imaginary. With real $\mu$, the saddle point of Section~\ref{sec:semiclassical}, at $\sinh^2\rho_\ast=\lambda/a\propto1/(c\mu)$, is then complex. A screen on a real slice requires continuing the coupling together with $c$, so that $c\mu$ stays real. The volume is then real, and the saddle point lies on a real slice at finite cosmological time, $\rho_\ast\to t_\ast/\ell$, in the spirit of holographic cosmology at finite time \cite{Araujo-Regado:2025ftc}. The two signs of the coupling are then replaced by a complex-conjugate pair. We leave this analysis to future work.

\subsection{Static patch}
\label{sec:static_patch}

As in the AdS$_3$ case (or conversely by reversing the map), the double
Wick rotation
\begin{align}
t\to i\ell\theta\ ,\qquad\qquad
\beta\to i\phi\ ,
\end{align}
maps the dS$_3$ torus universe to the Euclidean geometry
\begin{align}\label{EdS3torus}
ds^2
=
\ell^2\left(
d\theta^2
+\cos^2\theta\,d\alpha^2
+\sin^2\theta\,d\phi^2
\right),
\end{align}
which is the round three-sphere expressed in Hopf coordinates. The
coordinates range over $0\le\theta\le\pi/2$, while $\alpha$ and $\phi$
parametrize the two circle fibers of the Hopf fibration.

The hypersurface $\theta=\pi/4$ is the Clifford torus, where the two
Hopf fibers have equal radius. It is also the maximal-volume torus in
this foliation. Indeed, the trace of the extrinsic curvature of a
constant-$\theta$ hypersurface is
\begin{align}
\tau_E\equiv-K
=
\frac{2}{\ell}\cot(2\theta),
\end{align}
which vanishes at $\theta=\pi/4$.

Substituting this expression into the definition of the WdW time
\eqref{define_T} with $\Lambda=1/\ell^2$, we obtain the WdW time of this realization,
\begin{align}
T_H
=
\ln\cot\theta\ ,\qquad\mbox{or equivalently}\qquad \sinh T_H=\frac{\ell\tau_E}{2}\ .
\end{align}
As in Section~\ref{sec:ClosedU}, the additive constant in $T_H$ is fixed by
requiring the exponent of the kernel to be linear in the York time,
$\cosh(T_H-i\pi/2)=-i\ell\tau_E/2$, which is the shift induced by the double
Wick rotation, $\ln\coth(i\theta)=\ln\cot\theta-i\pi/2$. Thus $T=T_H-i\pi/2$:
the Clifford torus lies at $T=-i\pi/2$, on the vertical segment of $\mathcal C$,
while the CFT point $T=0$ lies off the real three-sphere and is reached by
continuing back to future infinity of the Lorentzian section. The Hopf $S^3$ is
thus the counterpart of the closed AdS$_3$ universe, with the roles of the two
signatures exchanged: in both cases, the realization with $\sinh T=\ell\tau/2$
contains the maximal-volume torus rather than the CFT point, and the Hagedorn
kernel on its real section is the pure phase $e^{2\lambda}e^{i\lambda\ell\tau_E}$.

The two branches differ here as in AdS$_3$, with the roles of the two signatures exchanged. On the Hagedorn branch, $\Gamma_+$ runs over the real Lorentzian slices, and the kernel is a wavepacket at future infinity; the three-sphere is reached only by shifting $\Gamma_+$ onto ${\rm Im}\,T=-\pi/2$. On the complex-energy branch, once its legs are moved onto ${\rm Im}\,T=\pm\pi/2$, the contour $\mathcal C$ starts at a pole of the three-sphere, where the $\phi$-circle shrinks smoothly, runs along the three-sphere to the Clifford torus, where $K=0$, and continues through the classically forbidden window to the Lorentzian universe at future infinity. This is the Hartle--Hawking no-boundary contour \cite{Hartle:1983ai}, now with a regular cap. Since the dS$_3$ torus universe has no $K=0$ slice, the Euclidean and Lorentzian geometries are connected through the classically forbidden window rather than glued directly. The correspondence between the two cases is summarized in Table~\ref{tab:AdS_dS}.

\begin{table}[!htb]
\centering\small
\setlength{\tabcolsep}{4pt}
\renewcommand{\arraystretch}{1.1}
\begin{tabular}{|l|c|c|}
\hline
 & \textbf{AdS$_3$} & \textbf{dS$_3$} \\
\hline
Real axis ($T=0$) & Euclidean AdS (boundary) & Lorentzian dS (future infinity) \\
Vertical segment & classically forbidden window & classically forbidden window \\
${\rm Im}\,T=-\pi/2$ & Lorentzian closed universe & Euclidean Hopf $S^3$ \\
$T=\pm i\pi/2$ ($K=0$) & maximal slice & Clifford torus \\
End of ${\rm Im}\,T=-\pi/2$ & big-bang singularity & pole of $S^3$ (regular cap) \\
\hline
Hagedorn: $\Gamma_+$ & Euclidean, shifted to Lorentzian & Lorentzian, shifted to $S^3$ \\
Complex energy: HH & boundary $\to K=0\to$ Lorentzian & cap $\to K=0\to$ Lorentzian \\
\hline
\end{tabular}
\caption{The complex $T$-plane for AdS$_3$ and dS$_3$ torus universes, with the roles of the two signatures exchanged. On the complex-energy branch, the deformed contour is a Hartle--Hawking contour, with the preparation starting at the asymptotic boundary for AdS$_3$ and at a regular cap for dS$_3$.}
\label{tab:AdS_dS}
\end{table}

This picture differs from the conventional Euclidean representation of
the dS/CFT correspondence \cite{Strominger:2001pn}, in which the
Lorentzian future (and past) infinity is mapped to the poles of the
three-sphere. Here, the CFT point remains at future infinity, while the
Clifford torus is distinguished as the maximal-volume torus at which the
complexified geometry connected to the CFT point meets the real three-sphere.

To relate this picture to the Lorentzian static patch, we introduce the
radial coordinate
\begin{align}
r=\ell\sin\theta,
\end{align}
which brings the metric into the form
\begin{align}
ds^2
=
\frac{dr^2}{1-r^2/\ell^2}
+\left(1-r^2/\ell^2\right)\ell^2d\alpha^2
+r^2d\phi^2.
\end{align}
Performing the Wick rotation
\begin{align}
\alpha\rightarrow it/\ell,
\end{align}
yields the Lorentzian static patch of dS$_3$. Under this map, the
Clifford torus, $\theta=\pi/4$, is mapped to the timelike surface
\begin{align}
r=\frac{\ell}{\sqrt2}.
\end{align}
The distinguished maximal-volume torus is therefore realized in the static
patch by the timelike surface $r=\ell/\sqrt2$.

Since our starting point is the
$T\bar{T}$-deformed torus partition function, this static-patch
description is understood as the analytic continuation of the Euclidean
Hopf $S^3$ with the periodic Euclidean time corresponding to the
Gibbons--Hawking temperature. The timelike surface $r=\ell/\sqrt2$ is
therefore naturally associated with a thermal family of static
observers. This is reminiscent of recent proposals emphasizing the
fundamental role of an observer in de Sitter quantum gravity based on
von Neumann algebras
\cite{Witten:2023qsv,Witten:2023xze,Chandrasekaran:2022cip}.
Although the underlying motivations are different, it is intriguing that
the holographic dictionary developed here likewise singles out a
preferred timelike surface --- and hence a preferred family of thermal
static observers --- as the locus at which the complex-energy contour meets
the real section.

\subsection{Flat Limit}
\label{sec:flat_limit}

The canonical quantization developed in
Section~\ref{sec:CQTUAdS} also applies to the case of vanishing
cosmological constant, $\Lambda=0$. The WdW equation retains the same
form as in \eqref{WdW_T}, while the relation between the WdW time $T$
and the York time is obtained from the $\Lambda\to0$ limit of
\eqref{define_T}. It is therefore natural to ask how the holographic
interpretation developed above extends to flat space.

The corresponding classical geometries are the Milne universe in
Lorentzian signature and the Rindler space in Euclidean signature,
\begin{equation}
\begin{aligned}
ds^2
&=
-dt^2+d\alpha^2+t^2d\beta^2
\qquad
\mbox{(Lorentzian)}\ ,
\\
ds^2
&=
dx^2+d\alpha^2+x^2d\phi^2
\qquad
\mbox{(Euclidean)}\ .
\end{aligned}
\end{equation}
Both geometries are locally flat and are related by the same double
Wick rotation as in the $(\mathrm{A})$dS cases. The Lorentzian geometry
is the Milne universe, while the Euclidean geometry is obtained by a
double Wick rotation of the Milne universe and has topology
$\mathbb{R}\times T^2$.
The coordinates range over $0\le t<\infty$ and $0\le x<\infty$. In the
Lorentzian geometry, $t=0$ corresponds to the Milne horizon, while
$t\to\infty$ approaches future timelike infinity. In the Euclidean
geometry, $x=0$ is the origin of the polar coordinates (the Euclidean
Rindler horizon), and $x\to\infty$ corresponds to spatial infinity.

The trace of the extrinsic curvature of a constant-$t$ (or constant-$x$)
hypersurface is
\begin{align}
\tau\equiv-K
=
\frac1t\ ,
\qquad\quad
\tau_E\equiv-K
=
\frac1x\ .
\end{align}
Substituting these expressions into the $\Lambda\to0$ limit of
\eqref{define_T}, we obtain
\begin{align}
T
=
-\ln\frac{t}{t_0}\ ,
\qquad\quad
T
=
-\ln\frac{x}{x_0}\ ,
\end{align}
for the Lorentzian and Euclidean geometries, respectively.

The absence of an intrinsic length scale implies that the reference
scales $t_0$ and $x_0$ are arbitrary. Consequently, the origin of the
WdW time, and hence the location of the corresponding ``holographic
screen,'' is not fixed a priori. By choosing the reference scales
appropriately, the screen may be placed on an arbitrary constant-$t$ (or
constant-$x$) hypersurface. In particular, taking
$t_0,x_0\rightarrow\infty$ moves the screen to spacetime infinity, where
it naturally coincides with the limiting maximal-volume slice, in
accordance with the $(\mathrm{A})$dS cases.


\section{Comments on Generalizations}
\label{sec:generalizations}

The present construction relies on several remarkable simplifications of the torus geometry. Nevertheless, it suggests a general framework for relating $T^2$-deformed field theories \cite{Taylor:2018xcy} to wavefunctions of quantum gravity in reduced phase-space quantization. A central challenge is therefore to understand the kernel that intertwines the field-theory and bulk descriptions.

Reduced phase-space quantization itself is not restricted to the torus, but extends naturally to arbitrary spatial topology and spacetime dimension. The principal obstacle is instead solving the Hamiltonian constraint. In general, the conformal factor is determined only implicitly through the Lichnerowicz equation \eqref{L_eqn}, making the reduced Hamiltonian a highly nonlocal functional of the reduced canonical variables. Consequently, the corresponding WdW equation is no longer expected to admit the simple separable form realized in the torus case.

From this perspective, the integral kernel introduced in the present work should be regarded as a fundamental object rather than merely a technical tool. In the torus case, it provides an explicit correspondence between the $T\Tb$ deformation parameter and the WdW evolution, realized as a change of representation from the York time to the volume, and leading to a simple holographic description in terms of finite radial or temporal resolution. Since the York time and the volume are canonically conjugate in any dimension, the part of the kernel implementing this change of representation is expected to persist in general; what changes is the Jacobian and the dependence on the remaining reduced phase-space variables. 
In more general settings, the kernel is expected to determine a much more complicated relation between the field-theory deformation scale and the bulk radial (or temporal) scale, potentially involving the additional geometric degrees of freedom of the reduced phase space.
Understanding the structure of such generalized kernels may therefore provide a natural framework for extending the present construction to punctured surfaces, higher-genus geometries, and higher-dimensional spacetimes.

More concretely, as reviewed in Section~\ref{sec:CQTUAdS}, the canonical
variables in reduced phase-space quantization are the conformal geometry
$q_A$ and its conjugate momentum $p^A$. The variables $q_A$ provide local
coordinates on the space of conformal geometries. Around a reference
conformal metric $\bar g^{(0)}$, one may locally write
\begin{equation}
\bar g_{ij}(x;q)
=
\bar g^{(0)}_{ij}(x)
+
q_A Y^{(A)}_{ij}(x)
+
O(q^2)\ ,
\end{equation}
where $Y^{(A)}_{ij}$ form a basis of transverse-traceless (TT)
deformations at $\bar g^{(0)}$. Only these TT deformations contribute to
the reduced symplectic form: the longitudinal degrees of freedom are
eliminated by the momentum constraint, while the trace degree of freedom
is traded for the York time through the Hamiltonian constraint. Their
conjugate momenta are
\begin{equation}
p^A
=
\int_\Sigma d^{d-1}x\,\sqrt{\bar g}\,
\bar\Sigma^{ij}
\frac{\partial\bar g_{ij}}{\partial q_A}\ ,
\end{equation}
where $\bar\Sigma^{ij}$ is the TT momentum conjugate to $\bar g_{ij}$.

The reduced Hamiltonian is the physical spatial volume,
\begin{equation}
H_{\rm red}(\tau,q,p)
=
V[\tau,\bar g(q),\bar\Sigma(q,p)]
=
\int_{\Sigma} d^{d-1}x\,\sqrt{g}\ ,
\end{equation}
where $g_{ij}=e^{2\phi}\bar g_{ij}$ and the conformal factor is determined
by the Hamiltonian constraint~\eqref{L_eqn}.

In the torus case, the WdW equation becomes separable, leading to the integral transform \eqref{forward_transform}
\begin{equation}
Z_{T\Tb}(\lambda,m_a)
=
\int_{\Gamma_\epsilon} dT\,
K_\epsilon(\lambda,T)\,
\Psi(T,m_a)\ ,
\end{equation}
where $\lambda=m_2/|\mu|$, and the kernel and the contour depend on the sign of $\mu$ as in \eqref{kernels}.
Motivated by the torus construction, we propose the following schematic generalization,
\begin{equation}
Z_{T^2}(\mu,q_0)
=
\int dT\,\mathcal D q\,
K(\mu,q_0;T,q)\,
\Psi(T,q)\ ,
\end{equation}
where $q_0$ denotes the conformal geometry on which the deformed field theory is defined, while $q$ labels the conformal geometry of the bulk wavefunction at WdW time $T$.
Unlike the torus case, where the kernel depends only on the dimensionless combination $\lambda=m_2/\mu$ and the WdW time $T$, the generalized kernel is expected to determine a much more complicated relation between the field-theory deformation scale and the bulk radial (or temporal) scale, potentially involving the additional geometric degrees of freedom of the reduced phase space.

Although the reduced Hamiltonian is generally unavailable in closed
form, it can in principle be constructed numerically by solving the
Lichnerowicz equation for fixed reduced canonical data. This may allow
the corresponding quantum dynamics in reduced phase-space quantization
to be studied numerically. Combined with a sufficiently detailed
understanding of the corresponding $T^2$-deformed partition function,
this may eventually make it possible to determine the generalized
integral kernel even when an analytic treatment is not available.


\section{Discussion and Outlook}
\label{sec:discussion}

\subsection*{Finite-resolution holography}

The central result of this work is that the holographic relation between the boundary theory and the bulk wavefunction is naturally formulated as an invertible integral transform rather than a pointwise identification. The transform is a change of representation from the York time to its conjugate, the spatial volume, so that a $T\bar T$-deformed partition function at fixed coupling is the bulk wavefunction at fixed volume, whose radial position is not sharp. The correspondence between field-theory and bulk scales therefore possesses a finite radial (or temporal) resolution, whose origin is the complementarity between the volume and the York time.

This interpretation also clarifies the role of the $T\bar T$ deformation. While the undeformed CFT partition function specifies the asymptotic boundary data, the family of $T\bar T$-deformed partition functions organizes the same information according to bulk scale. A partition function at fixed deformation parameter corresponds not to a sharply localized bulk slice but to a bulk wavepacket in York time at fixed volume. The complete bulk wavefunction is recovered only from the entire family of partition functions.

This viewpoint refines, rather than replaces, finite-cutoff holography \cite{McGough:2016lol,Kraus:2018xrn, Caputa:2020lpa}. For the semiclassical part of the bulk state, the transform localizes at large central charge on a finite cutoff surface, whose radius and quasi-local energy reproduce the finite-cutoff relation for the sign with complex energies. The cutoff surface is thus derived rather than assumed, and has a width of order $c^{-1/2}$. The normalizable part of the bulk state is subleading and defines no screen: below a threshold set by the deformation scale, its modes give the perturbative corrections through complex geometries in the classically forbidden window, and above it, for the sign with complex energies, the nonperturbative imaginary part of the partition function through Lorentzian slices behind the horizon.

\subsection*{The two signs of the coupling}

As summarized in Table~\ref{tab:branches}, the two signs of the coupling lead to sharply different screens. For the sign with Hagedorn growth, the transform is the projection of the Lorentzian Schr\"odinger wavefunction onto volume eigenstates, and its semiclassical screen is a glue-on surface beyond the asymptotic boundary~\cite{Apolo:2023vnm}, which ceases to exist where the deformed vacuum becomes complex. For the sign with complex energies, it is a Laplace transform through the classically forbidden window of York time, and its semiclassical screen is the finite cutoff surface. Both are captured by $r_c^2=-4G\ell/(\pi\mu)$, and are the two semiclassical limits of a single exact transform; the nonperturbative ambiguity left by the CFT limit is fixed by the positive-frequency condition of the reduced phase space. It would be interesting to compare this prescription with the nonperturbative structure found via resurgence \cite{Gu:2025tpy,Gu:2026rck}.

\subsection*{Relation to existing approaches}

The present construction is complementary to existing approaches to bulk reconstruction. In HKLL reconstruction \cite{Hamilton:2006az}, the radial coordinate determines the size of the boundary region over which local operators are smeared. Likewise, in entanglement wedge reconstruction
\cite{Czech:2012bh,Headrick:2014cta,Wall:2012uf,Jafferis:2015del,Dong:2016eik,Cotler:2017erl},
the size of the boundary subregion determines the depth of the corresponding entanglement wedge.

By contrast, the present construction concerns the holographic representation of global quantum states rather than local bulk observables. The relevant boundary scale is therefore determined not by the size of a boundary subregion but by the deformation scale together with the global geometry of the boundary torus. Within canonical approaches, the transform makes precise the relation between the volume and York-time representations considered in~\cite{Coleman:2020jte}, with the measure, contour, and inverse fixed by the reduced phase space.

\subsection*{Geometric realizations of the CFT limit}

\begin{table}[t]
\centering\small
\renewcommand{\arraystretch}{1.0}
\begin{tabular}{|l|c|c|}
\hline
\textbf{Bulk realization} &
\textbf{CFT point ($T=0$)} &
\textbf{Maximal-volume torus ($T=\mp i\pi/2$)} \\
\hline
Euclidean AdS$_3$
&
Asymptotic boundary
&
Not real (classically forbidden window) \\
\hline
AdS$_3$ closed torus universe
&
Off the real section
&
Maximal Cauchy slice \\
\hline
dS$_3$ torus universe
&
Future infinity
&
Not real (classically forbidden window) \\
\hline
Hopf $S^3$ / static patch of dS$_3$
&
Off the real section
&
Clifford torus / $r=\ell/\sqrt2$ \\
\hline
\end{tabular}
\caption{Geometric realizations of the CFT point $T=0$ and of the maximal-volume torus. The realizations related by a double Wick rotation differ by a shift of the WdW time by $i\pi/2$. Those with $\cosh T=\ell\tau/2$ contain the CFT point on their real section, while those with $\sinh T=\ell\tau/2$ contain the maximal-volume torus. The two loci are connected by the vertical segment of the contour $\mathcal C$ of the complex-energy branch, through the classically forbidden window $|\tau|<2/\ell$.}
\label{tab:CFTlimit}
\end{table}

Although the kernel transform takes the same mathematical form in all cases considered here, its geometric interpretation depends on the bulk realization. As summarized in Table~\ref{tab:CFTlimit}, the CFT point $T=0$ is the asymptotic boundary in Euclidean AdS$_3$ and future infinity in the dS$_3$ torus universe. In the double-Wick-rotated realizations, the closed AdS$_3$ universe and the Hopf three-sphere, the CFT point lies off the real section, which instead contains the maximal-volume torus, where the complexified geometry connected to the CFT point meets the real section. The corresponding observables are expected to depend on this geometric realization.

The de Sitter realizations are particularly suggestive from the perspective of de Sitter holography. For the de Sitter torus universe, the CFT limit is naturally realized at future infinity, consistent with the original proposal of de Sitter holography. More intriguingly, the static-patch realization singles out the timelike surface $r=\ell/\sqrt2$ associated with a static observer. This suggests that the holographic dictionary developed here may admit a natural observer-centered interpretation in de Sitter space. It would be interesting to understand the implications of this picture for observer-based formulations of de Sitter holography.

\appendix


\section*{Acknowledgments}

SH would like to thank Animik Ghosh, Rahul Poddar, and Masaki Shigemori for discussions and the department of mathematics at Nagoya University for their hospitalities during his visits where part of this work was done. 
SH also acknowledges the use of Claude Opus 5.5 (Anthropic) during the revision of this paper, for checking calculations, numerical verification, preparing figures, and editing the text. All results were verified by the author, who takes full responsibility for the content.


\end{document}